\documentclass[journal,letter,twoside]{IEEEtran}

\usepackage{cite}
\usepackage{amsmath,amssymb,amsfonts}
\usepackage{bm}
\usepackage{algorithmic}
\usepackage{textcomp}
\usepackage{comment}
\usepackage{verbatim}

\usepackage{setspace}

\def\smallspaceenum{}

\usepackage{floatrow}
\usepackage[caption=false,font=footnotesize]{subfig}

\usepackage[table, dvipsnames]{xcolor}  % Modern color package (used for text and tables)
\usepackage{fancyhdr}
\usepackage{graphicx}
\usepackage{url}
\usepackage{verbatim}
\usepackage{multirow}
\usepackage{hhline}
\usepackage[us]{datetime}
\usepackage{paralist}
\usepackage{color}
\usepackage{soul}
\usepackage{stfloats}
\usepackage[ruled, vlined, norelsize]{algorithm2e} % The algorithms writing package
\SetAlFnt{\sf} % Set the algorithms font style to sans serif

\graphicspath{{figures/}}

\usepackage{soul}

\newdimen\arrayruleHwidth
\makeatletter
\def\Hline{\noalign{\ifnum0=`}\fi\hrule \@height \arrayruleHwidth
\futurelet \@tempa\@xhline}
\makeatother

\newcolumntype{P}[1]{>{\centering\arraybackslash}p{#1}}

\makeatletter
\def\blfootnote{\xdef\@thefnmark{}\@footnotetext}
\makeatother

\shortdate
\settimeformat{ampmtime}

\ifCLASSINFOpdf
\else
\fi

\begin{document}

\newcommand{\thetitle}{Obfuscating the Interconnects: Low-Cost and Resilient Full-Chip Layout Camouflaging}

\title{\thetitle}

\author{Satwik~Patnaik,~\IEEEmembership{Student Member,~IEEE},
Mohammed~Ashraf,
Ozgur~Sinanoglu,~\IEEEmembership{Senior Member,~IEEE}, and
Johann~Knechtel,~\IEEEmembership{Member,~IEEE}
\thanks{%
This work is an extension of~\cite{patnaik17_Camo_BEOL_ICCAD}.
This work was supported in part by the Army Research Office under Grant 65513-CS, the Center for Cyber Security (CCS) at NYU/NYU AD, and the NYU
AD Research Enhancement Fund (REF) under Grant RE218.
The work of S.\ Patnaik was supported by the Global Ph.D. Fellowship at NYU/NYU AD.
Besides, this work was carried out in part on the HPC facility at NYU AD.
This paper was recommended by Associate Editor Y.\ Makris.
}
\thanks{S.\ Patnaik is with the Department
of Electrical and Computer Engineering, Tandon School of Engineering, New York University (NYU), Brooklyn, NY, 11201, USA (e-mail: sp4012@nyu.edu).} 
\thanks{M.\ Ashraf, O.\ Sinanoglu, and J.\ Knechtel are with the Division of Engineering, New York University Abu Dhabi (NYU AD), Saadiyat Island, 129188, UAE (e-mail: ma199@nyu.edu; ozgursin@nyu.edu; johann@nyu.edu).}
}

\maketitle

\renewcommand{\headrulewidth}{0.0pt}
\thispagestyle{fancy}
\lhead{}
\rhead{}
\chead{\copyright~2020 IEEE.
This is the author's version of the work. It is posted here for personal use.  Not for redistribution. The definitive Version of Record is
	published in IEEE TCAD, DOI 10.1109/TCAD.2020.2981034}
\cfoot{}

\begin{abstract}
Layout camouflaging
can protect the intellectual property of modern circuits.
Most prior art, however, incurs 
excessive layout overheads and
necessitates customization of active-device manufacturing processes, i.e., the front-end-of-line (FEOL).
As a result, camouflaging has typically been applied selectively, which can ultimately undermine its resilience.
Here, we propose a low-cost and generic scheme---full-chip camouflaging can be finally realized without reservations.
Our scheme is based on obfuscating the interconnects, i.e., the back-end-of-line (BEOL), through design-time handling for real and dummy wires and vias.
To that end, we implement custom,
BEOL-centric obfuscation cells, and develop a CAD flow using industrial tools.
Our scheme can be applied to any design and technology node without FEOL-level modifications. 
Considering its BEOL-centric nature, we advocate applying our scheme in conjunction with split manufacturing, to furthermore protect against
untrusted fabs.
We evaluate our scheme for various designs at the physical, DRC-clean layout level.
Our scheme incurs a significantly lower cost than most of the prior art. 
Notably, for fully camouflaged layouts,
we observe average power, performance, and area overheads of
24.96\%, 19.06\%, and 32.55\%, respectively.
We conduct a thorough security study addressing the threats (attacks) related to untrustworthy FEOL fabs (proximity attacks) and
malicious end-users (SAT-based attacks).
An empirical key finding is that
only large-scale camouflaging schemes like ours are
\textit{practically secure} against powerful SAT-based attacks.
Another key finding is that
our scheme
hinders both placement- and routing-centric proximity attacks;
correct connections are reduced by 7.47X, and complexity is increased by 24.15X, respectively, for such attacks.
\end{abstract}

\markboth{IEEE Transactions on Computer-Aided Design of Integrated Circuits and Systems}
{Patnaik \MakeLowercase{\textit{et al.}}: \thetitle}

\begin{IEEEkeywords}
Hardware security,
IP protection,
Reverse engineering, 
IC camouflaging,
Interconnects,
Split manufacturing,
Boolean satisfiability,
Proximity attacks.
\end{IEEEkeywords}

\section{Introduction}
\label{sec:introduction}

\IEEEPARstart{E}{nsuring} the integrity, security, and trustworthiness of integrated circuits (ICs) have become major concerns in recent
years~\cite{rostami14,TW12}.
One important aspect here is that the intellectual property (IP) of modern ICs can be duplicated without consent, resulting in financial losses
for IP vendors, which are estimated to amount to several billion dollars per year.
This is because modern ICs are the products of distributed and globalized supply chains. If an IC design is not protected, any of the various
entities involved in such
outsourced supply
chains may reconstruct/pirate the underlying IP.
Protecting the IP can also help
to mitigate other hardware-centric threats,
	e.g., hardware Trojans~\cite{TW12}.
		%BT18}.
Besides, a malicious end-user, i.e., an adversary without direct access to the design and fabrication process,
can resort to \textit{reverse engineering} of ICs to obtain the IP.
The tools and know-how for such reverse-engineering attacks are becoming more advanced and widely available, rendering this a practical
and severe threat~\cite{
	%torrance09,
	%skorobogatov12_chapter,
	TW12,
	sugawara14}.

\begin{figure}[tb]
\centering
\includegraphics[width=\columnwidth]{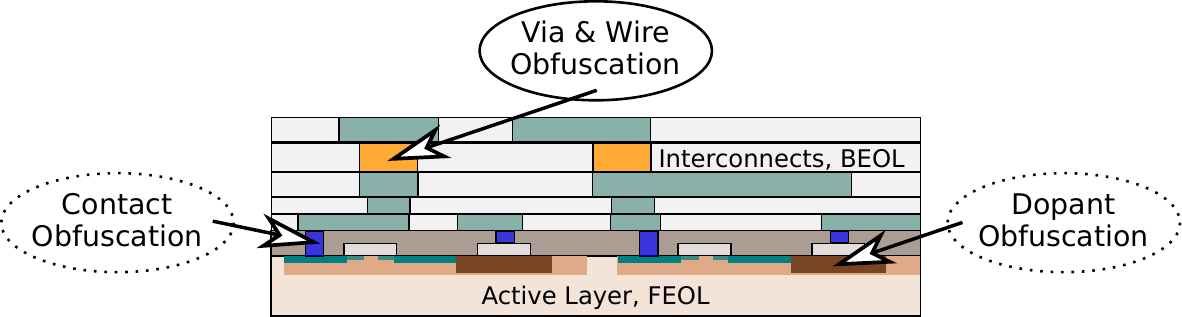}
\caption{Concepts for layout camouflaging. Most prior works target at the active layer, or front-end-of-line (FEOL), by stealthy manipulation of
gate contacts or dopant regions.
In contrast, our scheme is based on obfuscation of the interconnects, i.e., the back-end-of-line (BEOL).
\label{fig:LC}
}
\end{figure}

The traditional goal of \textit{layout camouflaging (LC)}, or simply \textit{camouflaging}, is to mitigate reverse-engineering attacks by end-users.
Therefore, camouflaging alters the physical appearance of an IC such that it is arduous 
for the attacker to
understand the true IC design (Fig.~\ref{fig:LC}).
In other words, when facing camouflaged ICs, an attacker shall be rendered unable to infer the chip IP directly, as the IP 
is hidden within an obfuscated physical design.\footnote{IP protection can also be realized at the system level, e.g., by
obfuscating finite state machines~\cite{lao15}. However, such techniques are orthogonal and independent from the physical obfuscation we consider in this work.}

Camouflaging is implemented traditionally at the device layer, e.g., by ``look-alike'' or ambiguous gates~\cite{cocchi14,rajendran13_camouflage}, by
secretly configured multiplexers~\cite{wang16_MUX,zhang16},
or by the threshold-dependent configuration of gates~\cite{collantes16,nirmala16,erbagci16}.
We note that prior art requires a fully trustworthy fabrication process, as
the device obfuscation has to be realized by the foundry.
We also note that most schemes incur a high layout cost.
For example, the XOR-NAND-NOR primitive proposed
in~\cite{rajendran13_camouflage} is expected to incur 4$\times$ area, 5.5$\times$ power, and 1.6$\times$ delay cost in comparison to a conventional NAND gate.
For another scheme by Akkaya \textit{et al.}~\cite{akkaya18}, demonstrated in a 65nm chip, an equivalent primitive
incurs even higher APD cost, namely
7.3$\times$ area (A), 9.2$\times$ power (P), and 6.6$\times$ delay (D).
Accordingly, the application of such schemes is suggested to be limited, which in turn may
compromise their resilience as discussed next.

Assuming that prior schemes are applicable only to selected parts of an IC in practice, an essential challenge is where and to what
extent camouflaging shall be effected.
Ideally,
an attacker's effort to resolve 
a carefully camouflaged netlist would be exponential in the number of camouflaged
gates~\cite{el2019sat}.
To achieve such resilience, among other approaches, Rajendran \textit{et al.}~\cite{rajendran13_camouflage} proposed to camouflage
gates that are interfering, i.e., gates that cannot be resolved individually.
Still, advances for \textit{Boolean satisfiability (SAT)} solvers have enabled
powerful attacks on camouflaging (and logic locking)~\cite{subramanyan15,yu17,el2019sat}, undermining the promises of~\cite{rajendran13_camouflage} and
other works.
Even for hardened schemes such as~\cite{li16_camouflaging, yasin16_CamoPerturb, xie16_SAT},
which explicitly aim for exponential attack resilience,
tailored SAT and other attack frameworks have been proposed~\cite{shamsi17,shen17,bypass-attack2017,
	%CycSAT-ICCAD2017,
	yasin17_TETC}.
Therefore, the trade-off between resilience and cost/applicability remains an open challenge for camouflaging.

In Table~\ref{tab:prior_camo_schemes}, we provide a high-level comparison of the selected prior art and our work.
We also discuss the prior art and other relevant security aspects further in Sec.~\ref{sec:background}.
Besides,
a comprehensive overview of camouflaging
is given in~\cite{vijayakumar16}, and IP protection in general
is reviewed in~\cite{rajendran14, knechtel19_IP_COINS}.

\begin{table}[tb]
\centering
\scriptsize
\setlength{\tabcolsep}{0.45mm}
\caption{Comparison of Selected Layout Camouflaging (LC) Schemes}
\label{tab:prior_camo_schemes}
\begin{tabular}{*{6}{c}}
\hline
\multirow{2}{*}{\textbf{Work}} & 
\textbf{LC} & 
\textbf{Trusted Foundry} & 
\textbf{LC} & 
\textbf{APD} & 
\textbf{Resilience to} \\
&
\textbf{Style} & 
\textbf{Assumed?} & 
\textbf{Scale} & 
\textbf{Cost} & 
\textbf{Algorithmic Attacks} \\
\hline

\cite{rajendran13_camouflage} 
& Contact 
& Yes
& ``Small''
& ``Very high''
& Vulnerable~\cite{el2019sat}, [This] \\ \hline

\cite{li16_camouflaging} 
& Contact/Dopant 
& Yes 
& ``Small''
& ``Moderate''
& Vulnerable~\cite{yasin17_TETC} \\ \hline

  \multirow{2}{*}{\cite{nirmala16}}
& \multirow{2}{*}{Dopant}
& \multirow{2}{*}{Yes}
& \multirow{2}{*}{``Small''}
& \multirow{2}{*}{``Very high''}
& Vulnerable for \\

&
&
&
&
& small-scale LC~[This] \\ \hline

\cite{chen18_interconnects} 
& {Interconnects}
& {Yes}
& {``Small''}
& {``Low''}
& {Vulnerable~[This]} \\ \hline

\multirow{2}{*}{\textit{Ours}}
& \multirow{2}{*}{\textit{Interconnects}}
& {\textit{Untrusted FEOL,}}
& \multirow{2}{*}{\textit{``Large''}}
& \multirow{2}{*}{\textit{``Moderate''}}
& \textit{Superior; attacks}
\\
& 
& {\textit{trusted BEOL}}
& 
& 
& \textit{yet to succeed} \\ \hline

\end{tabular}
\end{table}

\begin{figure}[tb]
\centering
\includegraphics[height=4.4em]{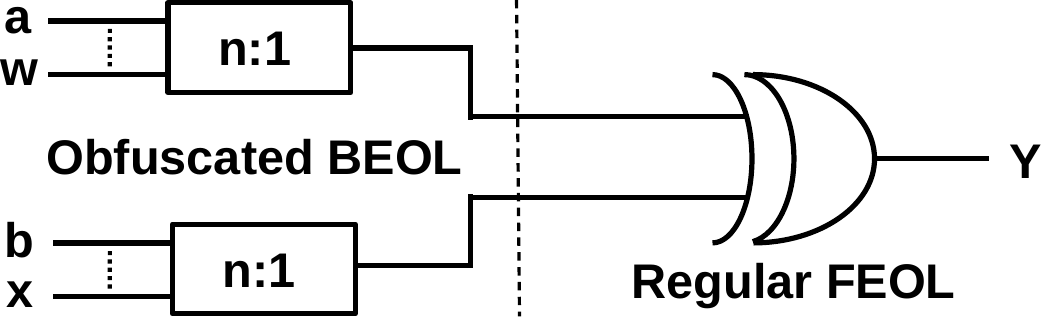}
\caption{Our concept is based on secret $n$:1 mappings at the BEOL, also using dummy vias and wires, thereby obfuscating the real inputs for any
	gate (not only two-input XOR gates).
The set of obfuscated functionalities
depends on $n$, the randomized selection of the dummy nets, and the gate type.}
\label{fig:concept}
\end{figure}

In this paper, we make the following case:
to remain resilient, at least as long as foreseeable, camouflaging should be applied at large scales, ideally for the \textit{full chip}.
Alternative approaches are required to enable
large-scale camouflaging, especially when aiming for low cost and fewer dependencies on trustworthy fabrication.
To that end, we propose and evaluate
a novel
concept based on \textit{obfuscating the interconnects} (Fig.~\ref{fig:concept}).
Our contributions can be summarized as follows:
\begin{compactenum}

\item
Based on the emerging notion of obfuscating the interconnects, we enable resilient and large-scale camouflaging, up to full-chip camouflaging. To do so, we develop a security-
and cost-driven, end-to-end CAD methodology, and
we propose novel back-end-of-line (BEOL)-centric camouflaging primitives which are applicable for all types of regular gates.
That is,
our primitives do not require any modification at the front-end-of-line (FEOL) layers and can, therefore, be easily integrated into any design, technology, and CAD flow.
We leverage \textit{Cadence Innovus}, and make our layouts publicly available in~\cite{webinterface}.

\item The fact that our work realizes BEOL-centric camouflaging suggests an application of \textit{split
	manufacturing}~\cite{mccants11,rostami14,vaidyanathan14_2}.
Doing so allows us, for the first time for camouflaging, to hinder fab adversaries in addition to malicious end-users.

\item We conduct a thorough evaluation of our scheme concerning the APD (or PPA---power, performance, and area)
cost for camouflaged, DRC-clean physical layouts.
In contrast, most
prior art
investigate their camouflaging primitives only as stand-alone devices, without applying them in actual layouts; we argue that this is overly optimistic.
Thus, our work is one of the very few
providing a comprehensive layout-level evaluation, and we also provide an in-depth and comprehensive study on APD cost for a wide range of prior camouflaging schemes.

\item We assess the resilience of our scheme
and compare it against previous works.
For our scheme, we introduce the notion of
\textit{practically secure camouflaging} which seeks to impose
high
computational costs on SAT-based attacks while, unlike \textit{provably secure camouflaging}, not requiring additional, dedicated circuit structures.
For that assessment, we employ powerful SAT-based attacks~\cite{code_pramod,shen17_code} on traditional benchmarks as well as large VLSI benchmarks.
With regards to split manufacturing, we employ state-of-the-art proximity attacks~\cite{code_network_flow_attack,code_MAGANA_attack} and demonstrate that our scheme is also resilient against
such fab-based adversarial activities.

\end{compactenum}

\section{Background and Motivation}
\label{sec:background}

\subsection{Split Manufacturing}
\label{sec:SM}

A technique orthogonal to camouflaging is \textit{split manufacturing}~\cite{mccants11,rostami14,vaidyanathan14_2}---while camouflaging seeks to protect against malicious
end-users, split manufacturing aims to hinder
adversaries acting at manufacturing time.
Typically, split manufacturing means to split the IC manufacturing flow into FEOL and BEOL processes, i.e., the front-end-of-line and back-end-of-line
steps, where the FEOL is outsourced, and the BEOL is realized by some trusted in-house or on-shore facility.
Note that split manufacturing has been demonstrated; \cite{vaidyanathan14_2}
describes promising measurement results for an \textit{IBM}/\textit{GlobalFoundries} 130nm split process, and \cite{mccants11}
summarizes further results,
most notably for a 28nm split process run by \textit{Samsung} across Austin
and South Korea.

For the FEOL foundry, the resulting partial layout appears as ``sea of gates,'' with most of the interconnects missing.
Therefore, it is argued that an adversary residing in the foundry cannot infer the full design easily, which hinders him/her from both IP
piracy and Trojan insertion.
The advantages of split manufacturing are as follows: 1) it allows for commissioning high-end, state-of-the-art, yet untrusted, FEOL facilities;
2) the BEOL is significantly less complex and cheap to implement compared to the FEOL and, thus, technology requirements for the trusted facility are
moderate;
3) the supply chain can remain largely as is, and additional steps are only required for preparation, shipping, and finalization of FEOL wafers.

Since physical layouts are designed holistically, at least when using regular and security-unaware CAD tools, various hints on the BEOL can
remain within the FEOL.
Wang \textit{et al.}~\cite{wang2018cat} proposed a so-called \textit{proximity attack}, where they formulated
various FEOL-level hints within a network-flow attack model.\footnote{The hints are:
(i)~physical proximity of gates,
(ii)~avoidance of combinatorial loops, which are rare in practice,
(iii)~timing and load constraints,
and (iv)~orientation and routing direction of wires.}
For selected designs, they succeeded to infer the majority of missing BEOL wires.
Maga\~{n}a \textit{et al.}~\cite{magana17} proposed various attacks that focus on FEOL-level routing,
and they conclude that such efforts
are more effective than relying solely on the placement.
Zhang~\textit{et al.}~\cite{zhang18}
and Li~\textit{et al.}~\cite{li19_SM_ML_DAC}
leveraged machine learning for advanced and scalable attack frameworks.
However, neither of those attacks~\cite{zhang18,magana17,li19_SM_ML_DAC} recovers an actual netlist; they instead provide sets of
most probable BEOL connections. Therefore, an attacker requires further effort and know-how
before he/she can obtain a
complete netlist.
In any case, attacks which can achieve 100\% correctness when inferring all BEOL wires of industrial, large-scale designs, possibly even with
some placement- or routing-level perturbations introduced for
protection (e.g.,
see~\cite{
sengupta17_SM_ICCAD,
feng17,patnaik18_SM_ASPDAC,patnaik18_SM_DAC}),
are yet to be demonstrated.

\begin{table*}[tb]
\centering
\scriptsize
\setlength{\tabcolsep}{0.85mm}
\caption{Post-Layout Results on selected \textit{ITC-99} Benchmarks
and Resulting Camouflaging Limits for Selected Schemes
}
\label{tab:lib_dependency_camo}
\begin{tabular}{cccccccccccccccccccc}
\hline
\multirow{3}{*}{\textbf{Benchmark}}
&
& \multirow{2}{*}{\textbf{Total}}
&
& \multicolumn{10}{c}{\textbf{Instance Counts of Relevant Cells}}
&
& \multicolumn{5}{c}{\textbf{Resulting Camouflaging Limit [\% of Total Instances]}} \\
\cline{5-14}
\cline{16-20}

&
& \multirow{2}{*}{\textbf{Instances}}
&
& \multirow{2}{*}{\textbf{XOR2}} & \multirow{2}{*}{\textbf{XNOR2}}
& \multirow{2}{*}{\textbf{OR2}} &  \multirow{2}{*}{\textbf{NOR2}}
& \multirow{2}{*}{\textbf{NAND2}} &\multirow{2}{*}{\textbf{AND2}}
& \multirow{2}{*}{\textbf{INV}} &  \multirow{2}{*}{\textbf{BUF}}
& \multirow{2}{*}{\textbf{AND3}} & \multirow{2}{*}{\textbf{NAND3}}
& 
& \textbf{XOR-type~}
& \textbf{STF-type}
& \textbf{XOR-NAND-NOR}
& \textbf{Threshold}
& \multirow{2}{*}{\textbf{Ours}}
\\

&
&
&
& & 
& & 
& & 
& & 
& & 
&
& \textbf{\cite{li16_camouflaging}}
& \textbf{\cite{li16_camouflaging}}
& \textbf{\cite{rajendran13_camouflage}}
& \textbf{\cite{nirmala16}}
& 
\\ \hline 

b14\_C
&
& 3,263
&
& 11 & 143 & 144 & 156 & 930 & 248 & 636 & 23 & 22 & 89
&
& 68.89 
& 58.57
& 33.62
& 50.01
& 100
\\ %\hline

b15\_C 
&
& 4,972
&
& 16 & 61 & 133 & 263 & 1,285 & 321 & 844 & 156 & 24 & 128
&
& 63.44 
& 54.4
& 31.46
& 41.81
& 100
\\ %\hline

b17\_C
&
& 16,169
&
& 34 & 207 & 553 & 1,012 & 3,561 & 876 & 3,143 & 536 & 41 & 465
&
& 63.01 
& 54.71
& 28.49
& 38.61
& 100
\\ %\hline

b20\_C 
&
& 7,184
&
& 40 & 288 & 318 & 355 & 2,087 & 451 & 1,489 & 117 & 49 & 263
&
& 71.39 
& 61.46
& 34.55
& 49.26
& 100
\\ %\hline

b21\_C 
&
& 7,318
&
& 33 & 341 & 407 & 401 & 1,949 & 551 & 1,505 & 110 & 41 & 206
&
& 70.65 
& 60.3
& 32.56
& 50.31
& 100
\\ %\hline

b22\_C 
&
& 10,839
&
& 72 & 457 & 544 & 595 & 3,058 & 653 & 2,194 & 170 & 60 & 379 
&
& 70.61 
& 61.09
& 34.37
& 49.63
& 100
\\ \hline

\textbf{Average} 
&
& --
&
& -- & -- & -- & -- & -- & -- & -- & -- & -- & --
&
& 67.99 
& 58.42
& 32.51
& 46.61
& 100
\\ \hline

\end{tabular}
\\[1mm]
We use the \textit{NanGate} 45nm library~\cite{nangate11}. All benchmarks are set up for iso-performance at 5ns.
The resulting \textit{camouflaging limit} (right)
is the ratio of cells that can be camouflaged over total cell instances, the latter as obtained by regular
synthesis using the full library (left).
Cells described as relevant for XOR-type camouflaging in \cite{li16_camouflaging}: BUF, INV, AND2, NAND2, OR2, NOR2, AND3, NAND3;
cells described as relevant for STF-type camouflaging in \cite{li16_camouflaging}: BUF, INV, AND2, NAND2, OR2, AND3;
cells assumed relevant for XOR-NAND-NOR camouflaging \cite{rajendran13_camouflage}: XOR2, NAND2, NOR2;
cells assumed relevant for threshold-dependent camouflaging \cite{nirmala16}: AND2, NAND2, NOR2, OR2, XOR2, XNOR2.
\end{table*}

\subsection{Camouflaging at the FEOL: Limitations and Vulnerabilities}
\label{sec:LC_FEOL}

Prior art
targets mainly at the active device layer, the FEOL,
and, thus, requires (i)~custom design of camouflaged gates, along with library characterization, and
(ii)~some modifications of well-established and optimized device manufacturing processes.
It is easy to see that both will incur efforts and financial cost, that is on top of any layout overheads.

Regarding layout overheads, as already indicated in Sec.~\ref{sec:introduction}, most existing camouflaging schemes incur a high cost
and are accordingly limited for practical use. 
For example for Zhang's work~\cite{zhang16},
we observe 464\%, 638\%, and 63\% overheads for
area, power, and delay,
	respectively, when camouflaging all gates of the \textit{ISCAS-85} benchmark \textit{c7552};
these numbers are 13.92$\times$, 37.51$\times$, and 4.47$\times$ higher than ours for the same scenario (see also
Sec.~\ref{sec:layout_evaluation} for more comparisons).
	
Furthermore, it is important to note that prior art can be constrained to begin with, namely by \textit{synthesis} limiting
the extent or scale to which camouflaging is
applicable at all.
More specifically, most prior arts 
implement stealthily multiple Boolean functions within customized gate-level structures, but these primitives
capture typically only some subsets of all the functionalities needed for modern IC designs.
For example, with the XOR2-NAND2-NOR2 primitive proposed in~\cite{rajendran13_camouflage},
the \textit{camouflaging scale} is inherently constrained (by synthesis) to the ratio of gates being mapped to
XOR2, NAND2, and NOR2.\footnote{For FEOL-centric camouflaging, the camouflaging scale would be defined as the ratio of camouflaged gates
over the total number of gates in the layout. This definition applies for this particular section, and for the
discussion of the prior art throughout this paper.
For our BEOL-centric scheme, however, camouflaging scale is defined as the number of obfuscated nets
over the total number of nets.
It is important to note that these two definitions are compatible; that is because we always obfuscate all the inputs of any
gate selected for camouflaging.}
We explore this limitation for selected schemes in Table~\ref{tab:lib_dependency_camo}.
We note that the limited scales of around 30--40\% camouflaging observed for some schemes are \textit{not} sufficient to protect against SAT-based attacks; see
also Sec.~\ref{sec:security_evaluation_SAT}
for details.
We also note that, as a countermeasure,
one may conduct synthesis with constrained libraries that only provide the functions covered by the employed camouflaging primitives.
Doing so, however, would incur considerable APD cost over full-library synthesis already for the baseline
(Table~\ref{tab:initial_overheads}), with actual overheads incurred by the
camouflaging primitives themselves coming further on top.

\begin{table}[tb]
\centering
\scriptsize
\setlength{\tabcolsep}{0.88mm}
\caption{Post-Layout Area (A), Power (P), and Delay (D) Baseline Cost, \textit{Without Camouflaging}, but for Constrained Synthesis}
\label{tab:initial_overheads}
\begin{tabular}{*{13}{c}}
\hline
\multirow{2}{*}{\textbf{Benchmark}}
&
& \multicolumn{3}{c}
{\textbf{XOR-type~\cite{li16_camouflaging}}} 
&
& \multicolumn{3}{c}
{\textbf{XOR-NAND-NOR~\cite{rajendran13_camouflage}}}
&
& \multicolumn{3}{c}
{\textbf{Threshold~\cite{nirmala16}}}
\\
\cline{3-5}
\cline{7-9}
\cline{11-13}

&
& \textbf{A} & \textbf{P} & \textbf{D} 
&
& \textbf{A} & \textbf{P} & \textbf{D} 
&
& \textbf{A} & \textbf{P} & \textbf{D}   
\\
\hline

b14\_C & 
&
5.63 & -7.25 & -1.85 &
&
18.24 & 8.81 & -0.27 &
&
13.24 & 1.21 & -0.27 
\\ %\hline

b15\_C & 
&
24.99 & 11.79 & -1.06 &
&
39.05 & 31.93 & -0.79 &
&
32.35 & 22.79 & -1.99 
\\ %\hline

b17\_C & 
&
28.38 & 20.13 & 0.99 &
&
33.96 & 21.81 & -1.25 &
&
38.18 & 25.09 & 4.92 
\\ %\hline

b20\_C & 
&
14.14 & -1.38 & 0.64 &
&
25.62 & 14.97 & -1.07 &
&
18.39 & 7.76 & 0.04 
\\ %\hline

b21\_C & 
&
17.25 & 3.11 & 0.27 &
&
23.11 & 8.86 & 0.76 &
&
16.19 & -0.06 & 0.37 
\\ %\hline

b22\_C & 
&
16.29 & 5.24 & -0.79 &
&
29.88 & 25.32 & 0.49 &
&
19.25 & 10.71 & 0.21 
\\ \hline

\bf{Average} & 
&
17.78 & 5.27 & -0.3 &
&
28.31 & 18.62 & -0.36 &
&
22.93 & 11.25 & 0.55 
\\
\hline
\end{tabular}
\\[1mm]
All cost are in percentage.
We use the \textit{NanGate} 45nm library~\cite{nangate11}, with synthesis runs
constrained individually to cells relevant
for each camouflaging scheme (see Table~\ref{tab:lib_dependency_camo}).
For a fair comparison, all
runs
are configured for iso-performance at 5ns.
\end{table}

In short, existing schemes can be applied only selectively---if at all---due to their
impact on FEOL processing steps, their
inherent APD overheads, and their limited
scale.
As a result, the constrained application of these and other schemes can compromise their resilience, as discussed next.

As mentioned in Sec.~\ref{sec:introduction}, powerful SAT-based attacks have challenged most prior art on camouflaging (and logic locking)~\cite{subramanyan15,yu17,el2019sat}.
The idea is that an obfuscated netlist can be modeled as satisfiability problem, where a working chip copy is leveraged as an
\textit{oracle} to
iteratively revise the model of the camouflaged netlist, by pruning out step by step any infeasible assignments for the camouflaged gates.
The previously unforeseen success of such SAT-based attacks stems from the fact that typically only a small number of I/O patterns are
required when resolving various camouflaging schemes.

Several studies on \textit{provably secure} camouflaging (and logic locking), e.g., see~\cite{li16_camouflaging, yasin16_CamoPerturb,xie16_SAT,yasin17_CCS}, seek to mitigate
SAT-based attacks by inserting dedicated
structures which necessitate considering an
exponential number of I/O patterns.
However, the output signals related to such structures 
have been identified successfully, e.g., by signal-probability- or sensitization-driven algorithms~\cite{yasin17_TETC,jiang2018efficient,yasin17_Anti-SAT}.
These critical wires may be ignored or even cut to circumvent the security features. We note that
wire cutting has been demonstrated in the past~\cite{helfmeier13}, enabling such invasive attacks in principle.
Besides, these schemes are also vulnerable to
other algorithmic attacks~\cite{shamsi17,shen17,bypass-attack2017,
	%CycSAT-ICCAD2017,
	yasin17_TETC}.
This is because a key limitation of these schemes is that they induce low output
corruptibility,
allowing an attacker to obtain relatively easily an \textit{approximate} version of the IP.
Note that this fact also implies that
these schemes are less suitable to protect design IP at the chip scale.

Reverse engineering measures may also render FEOL-centric camouflaging directly void, without the assistance of algorithmic techniques.
Schemes such as ``look-alike'' and ambiguous gates~\cite{cocchi14,rajendran13_camouflage}, or secretly configured MUXes~\cite{wang16_MUX,zhang16} rely on dummy contacts or dummy channels.
While yet to be demonstrated,
we argue that scanning electron microscopy in the passive voltage mode (SEM PVC) might break these schemes.
This is because dummy contacts/channels accumulate charges to a much lower degree than real contacts/channels.
Threshold-dependent camouflaging of gates
can also be revealed by SEM PVC,
as demonstrated successfully
by Sugawara \textit{et al.}~\cite{sugawara14}.
As Collantes \textit{et al.}~\cite{collantes16} indicate themselves, monitoring the etch rates can also reveal different doping levels, which are at the heart of
their threshold-dependent gates.
Besides, the tailored structures of provably secure schemes may be relatively easy to identify during reverse engineering, as these structures are typically
applied only in few places due to their relatively high cost, and they rely on
otherwise uncommon, very large combinatorial trees.

Overall, considering these various limitations and vulnerabilities
for prior art on FEOL-centric camouflaging, we argue that alternative
approaches are necessary.
We believe that such novel approaches should foremost be large-scale camouflaging schemes, ideally camouflaging full chips, to 1)~remain resilient as long as
foreseeable against any algorithmic attacks, and 2)~impose the highest efforts for any physical attacks.

\subsection{Toward Camouflaging at the BEOL}
\label{sec:LC_BEOL}

An interesting approach was suggested by Chen \textit{et al.}~\cite{chen18_interconnects}, namely to obfuscate the interconnects by implementing
real, conductive vias using magnesium (Mg) along with dummy, non-conductive vias using magnesium oxide (MgO).
Chen \textit{et al.\ }elaborated
that the use of Mg/MgO
is practical from both the perspectives of (i)~manufacturability and (ii)~mitigation of reverse engineering. For~(i),
they noted
that Mg had been used traditionally to facilitate the bonding of copper interconnects,
     and both Mg/MgO are compatible with
standard processes, in particular sputtering, electron-beam evaporation, and Dual-Damascene.
Independently, Swerts \textit{et al.}~\cite{swerts15_Mg_BEOL,swerts17} and Hwang \textit{et al.}~\cite{hwang12_transient_electronics} have used Mg and MgO for
customized BEOL processes, albeit without camouflaging in mind.
For~(ii), Chen \textit{et al.\ }fabricated samples and observed that Mg was completely oxidized into MgO
within a few minutes.
That is, the real Mg vias became indistinguishable from the dummy MgO vias during reverse engineering.
Again independently, Hwang \textit{et al.}~\cite{hwang12_transient_electronics} have shown that Mg not only oxidizes but also dissolves
quickly---as does MgO---when surrounded by fluids. We note that such dissolving would be inevitable
for etching and other reverse-engineering procedures.

Although one can argue that reverse engineering of Mg/MgO vias is possible nevertheless, this is yet to be demonstrated.
In general, the notion of
camouflaging at the BEOL
is not limited to
materials currently established for manufacturing;
future interconnects, e.g., based on
carbon/graphene or spintronics~\cite{CNT_TSDM17, naeemi14}, may become relevant as well.

\begin{figure}[tb]
\centering
\includegraphics[width=\columnwidth]{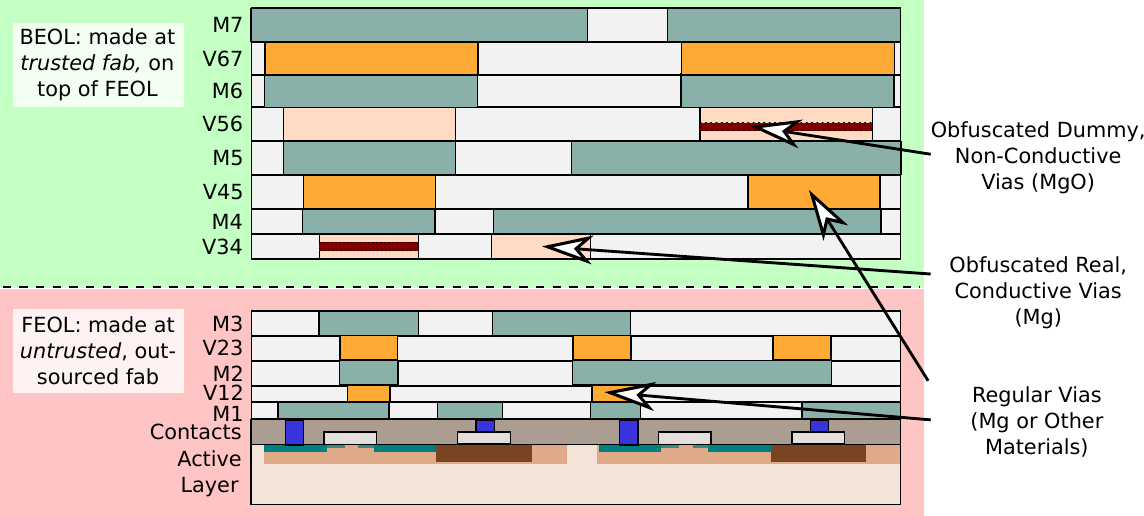}
\caption{Camouflaging at the BEOL implemented along with split manufacturing---this scheme helps to protect against threats arising from both fab adversaries and malicious
	end-users.
	True, conductive magnesium (Mg) vias are used in conjunction with dummy, non-conductive magnesium
	oxide (MgO) vias. During reverse engineering, the true Mg vias oxidize into MgO vias, rendering the differentiation of true and dummy
		vias difficult~\cite{chen18_interconnects}.}
\label{fig:LC_BEOL}
\end{figure}

It is important to
note that
obfuscated interconnects, similar to FEOL-centric camouflaging,
are implemented not only at the security-enforcing designer's choice but also at the manufacturer's discretion.
However, obfuscating the interconnects still allows to limit the dependency on outsourced, potentially untrustworthy manufacturing facilities,
since only the less complex BEOL part
has to remain protected within trusted facilities. We note that split manufacturing is particularly promising in this context
(Fig.~\ref{fig:LC_BEOL}).

Finally, another argument for obfuscating the interconnects is that commercial cost may be compensated for, even when
split manufacturing is  applied.
This is because one requires only additional BEOL processing steps, with BEOL masks being relatively cheap; e.g., for 16nm, M5/M6 masks are 3.5--4$\times$ cheaper 
than Poly masks~\cite{mask_cost}.
In contrast, most FEOL-centric camouflaging
(but not necessarily threshold-dependent camouflaging)
incurs high cost for the additional FEOL masks.
In general, we would like to recall that FEOL-centric camouflaging demands some alterations for devices and their manufacturing process---such alterations are likely
more complex and costly than camouflaging at the BEOL.

Despite their pioneering work,
Chen \textit{et al.\ }did
not provide
an
SAT-resilient scheme; see Sec.~\ref{sec:security_evaluation_SAT}.
We also emphasize that our concept and methodology differ significantly from~\cite{
		chen18_interconnects}; here, we only leverage their
notion of Mg/MgO vias.
Moreover, we advocate and enable the obfuscation of interconnects for large-scale, even full-chip camouflaging, by proposing dedicated
design-time techniques to handle dummy vias and wires for any gate.
In contrast, 
Chen \textit{et al.}\cite{chen18_interconnects} considered only small-scale, selected use of additional
dummy wires/vias.

\section{Concept and Camouflaging Primitive}
\label{sec:concept}

\begin{figure}[tb]
\centering
\includegraphics[scale=.433]{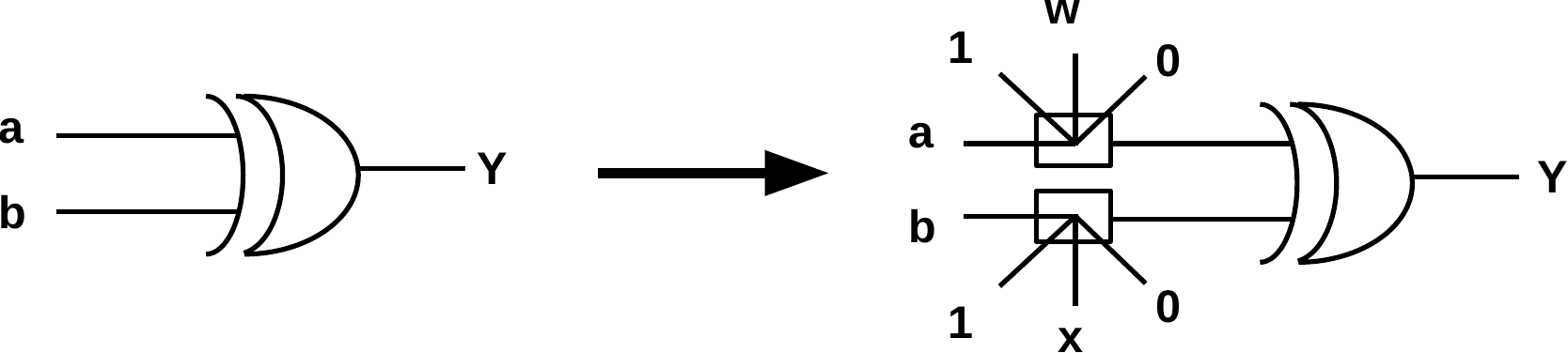}
\caption{Our final primitive, with fixed logic values \textit{0} and \textit{1}
along with the real and randomly selected dummy nets.
Depending on the gate type, 10 or 14 obfuscated functionalities arise for two-input gates.
For this example
with XOR as the underlying gate,
we can obfuscate the following 14 functionalities:\vspace{0.5ex}
$\textit{0},\textit{1},a,b,w,x,\overline{a},\overline{b},\overline{w},\overline{x}$,
	$a \oplus b$,
	$a \oplus x$,
	$w \oplus b$, and
	$w \oplus x$.
Recall that our concept is generic and directly applicable for any multi-input gates, not only for two-input gates.
}
\label{fig:final_primitive}
\end{figure}

Our key idea is the following:
we leverage reverse-engineering-resilient and obfuscated interconnects
to design novel, BEOL-centric camouflaging primitives that apply to any regular gate (Fig.~\ref{fig:concept}).
We implement $n$ wires
for each input of the gates to be camouflaged, where $n-1$ of those wires are randomly selected to be acting as dummies, and all $n$ wires have
their vias obfuscated.
In other words,
we obfuscate the real inputs of any gate via \textit{secret and randomized $n$:1 mapping at the BEOL}.
As a result, the actual function which a gate implements is obscured in a simple yet effective manner.
The set of possible functionalities
depends on $n$, the selection of the $n$ wires, and on the gate itself; a representative example is given in Fig.~\ref{fig:final_primitive}.

We conducted an exploratory, yet thorough, study on different flavors of our camouflaging primitive.
We assessed all flavors when applied at large scales for various benchmarks by 1)~their APD cost, while taking our scheme end-to-end and
evaluating GDSII layouts, and 2)~their
resilience against SAT-based attacks while running the seminal attack~\cite{subramanyan15,code_pramod}.
The exploratory study is detailed in~\cite{patnaik17_Camo_BEOL_ICCAD}; we summarize it next.

We first explored the most basic flavor with two wires for each gate's input, i.e., one randomly selected dummy wire and one real wire.
We observed that this flavor offers relatively weak resilience against SAT-based attacks, even for large-scale camouflaging.
Therefore, we next extended our primitive for
two dummy wires,
    which provided notably stronger resilience.
However, we also noted considerable APD cost for that flavor, due to the following.
To hinder an attacker from identifying dummy wires as such, we have to select all the nets for dummy wires randomly but also such that there are no combinatorial loops.
Furthermore, dummy wires have to be driven by unique nets, as otherwise some of the possible functionalities could be ruled out directly.
These requirements translated to significantly increased routing congestion in practice and, in turn, higher power consumption and delays.
Therefore, we note that the use of dummy wires should be limited for large-scale applicability of our scheme.
Finally, we explored an extended flavor comprising 
fixed logic values \textit{0} and \textit{1} along with regular nets
(Fig.~\ref{fig:final_primitive}).
When compared to employing only regular nets,
the benefit of using fixed-value nets is that the latter 
are not switching, thus
exhibiting only negligible power consumption and imposing no timing overhead.
Also, many of these wires can have the same driver
(but not all, see Sec.~\ref{sec:fixed_values}),
which helped to reduce the routing
congestion.
Regarding SAT-based attacks, we observed that this flavor provides good resilience.

In short, we select the flavor exemplified in Fig.~\ref{fig:final_primitive} for the remainder of our study, as we found empirically that it offers the best security-cost trade-off~\cite{patnaik17_Camo_BEOL_ICCAD}.

Since our concept is based on camouflaging at the BEOL,
it can be readily applied in conjunction with split manufacturing
(Sec.~\ref{sec:SM}).
To that end, we tailor our camouflaging primitive initially for higher metal layers, namely for M5 and M6 (Sec.~\ref{sec:obfuscating_cell}).
We
do so because split manufacturing is considered more practical for higher layers~\cite{xiao15,patnaik18_SM_ASPDAC}, as those layers have rather large pitches,
which are relatively easy and cheap to manufacture by the BEOL facility.
We also explore our camouflaging primitive when tailored for
lower metal layers (i.e., M3 and M4 in this work),
      and we support both types of primitives holistically in our flow (Sec.~\ref{sec:CAD_flow}).

\section{Threat Model}
\label{sec:threat_model}

We assume both the end-users and the fab to be untrusted.
The latter is in direct contrast to
the prior art on camouflaging that \textit{has to} trust the fab because of their FEOL-centric techniques.
To hinder fab adversaries, i.e., to protect the secret dummy/true vias and wires in the BEOL layers, we leverage split manufacturing;
therefore, we require a trusted BEOL facility.\footnote{
	In a weaker threat model where the fab is \textit{trusted},
our scheme and its primitives can still be applied as is, just \textit{without} need for split manufacturing.}
The goal of malicious end-users is to reverse engineer the chip's layout and identify its secret mappings in the BEOL layers---recall that the latter is challenging and yet to be demonstrated
(Sec.~\ref{sec:LC_BEOL}).
Ultimately, both malicious end-users and fab adversaries want to reconstruct the original netlist and its IP.
To that end,
we assume that end-users employ SAT-based attacks, whereas fab workers launch proximity attacks.
We evaluate both threat scenarios in Sec.~\ref{sec:security_evaluation_SAT} and \ref{sec:security_evaluation_PA}.

We assume that malicious end-users can get hold of working chips, which are used
(i)~as oracle for iterative SAT-based attack, and
(ii)~for reverse engineering of the chip's layout (as required to model the obfuscated netlist as an SAT problem), but end-users cannot directly
infer the secret BEOL mappings.
We assume that malicious fab workers have access to the FEOL layers, but not to the BEOL layers.
We further assume that both end-users and fab workers have the
complete know-how of our scheme, the used design tools, etc.
However, the adversaries cannot, by nature, reproduce the randomized selection steps taken throughout our scheme.

\section{Camouflaging Methodology}
\label{sec:methodology}

\subsection{Protecting Fixed Values and ``Implausible Functionalities''}
\label{sec:fixed_values}

For large-scale camouflaging using our primitive (Fig.~\ref{fig:final_primitive}), we note that the ubiquitous wires relating to fixed values \textit{0} and \textit{1}
may give away clues to an attacker.
Such fixed-value wires would be driven by distinct, easy-to-identify \textit{TIE cells}, and 
such cells and wires are typically used only selectively, e.g., for special registers or ``hardware feature flags.''
Therefore, an attacker observing a vast number of fixed-value wires
might rightfully
assume that these wires have been introduced only for obfuscation, and could try to disregard them.
Besides, we have to address another
challenge at once:
a mindful attacker may also try to rule out
all the ``implausible functionalities''
which are those beyond any gate's original functionality,
i.e., inverter, buffer, and TIE cells,
from the search space.
Since these functionalities arise only due to the fixed values being part of the obfuscated inputs, they
can only become effective once the fixed values are made an essential part of the design, and vice versa.

In short, the fixed-value wires have to be
rendered essential parts of the layout, while also protecting all the ``implausible functionalities,'' of gates.
To do so, we perform 
\textit{netlist transformations} at the beginning of our methodology as follows:

\begin{figure}[tb]
\centering
\includegraphics[scale=.433]{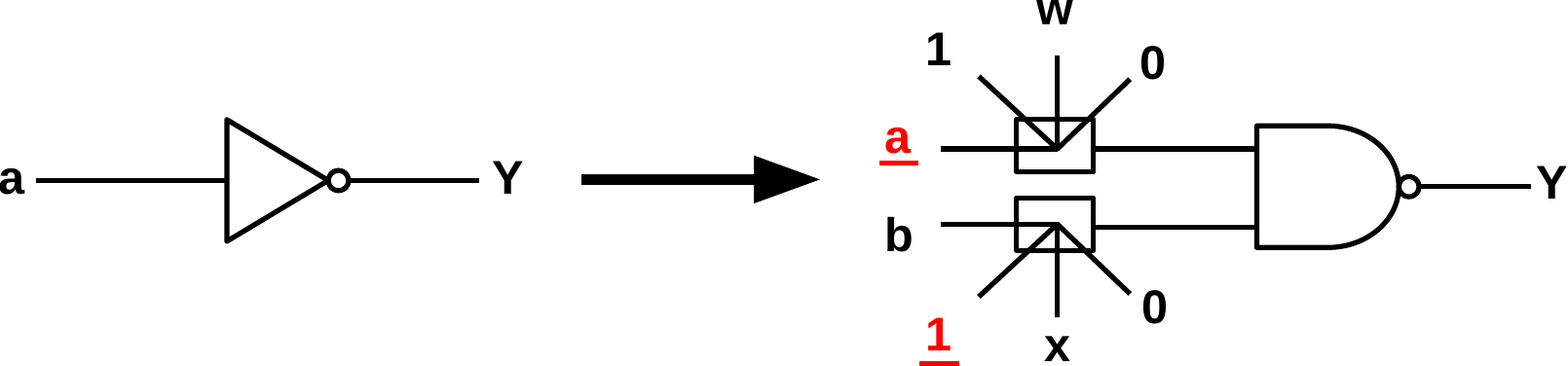}
\caption{An inverter transformed into a two-input NAND gate.
The real nets/wires are underlined and shown in red.
}
\label{fig:primitive_inv}
\end{figure}

\smallspaceenum
\begin{compactenum}

\item We transform selected inverters (INVs) and buffers (BUFs) into
camouflaged gates of other types (e.g., see Fig.~\ref{fig:primitive_inv}).
Nowadays, around 50\% of all gates are repeaters using INV/BUF~\cite{
		%shelar13,
	markov14}, offering ample opportunities for related
transformations.
One can freely choose (i)~the number of INVs/BUFs to transform and (ii)~ the type of gate to transform them into following Boolean algebra.
For a standard library, for example,
INV (BUF) can
be implemented using NAND (AND), NOR (OR), as well as both XOR and XNOR, all while using the fixed-value wires as one of the real inputs for
those gates.
This step renders the fixed-value wires essential---they cannot be ignored anymore without misinterpreting those
transformed, camouflaged gates and, therefore, without misinterpreting the whole design.
We also note that the ``implausible functionalities'' INV and BUF are now rendered plausible for the design and, independently,
an attacker cannot simply infer a direct relationship between any of the camouflaged gates and their true functionality.

\item We insert additional gates of randomly selected types into randomly selected regions of whitespace. These additional gates are camouflaged
as well, with their real inputs tied to fixed-value wires.
These gates act as ``TIE cells in disguise'' and, in turn, they
seemingly ``drive'' other camouflaged gates, which
are randomly selected from the vicinity.
Therefore, also the ``implausible functionalities'' \textit{0} and \textit{1} cannot be ruled out anymore without misinterpreting the design.
As with the transformation of INV/BUF above, leveraging randomness here is essential to hinder attackers from simply inferring these camouflaged gates.

\end{compactenum}

\subsection{CAD Flow}
\label{sec:CAD_flow}

\begin{figure}[tb]
\centering
\includegraphics[width=.75\textwidth]{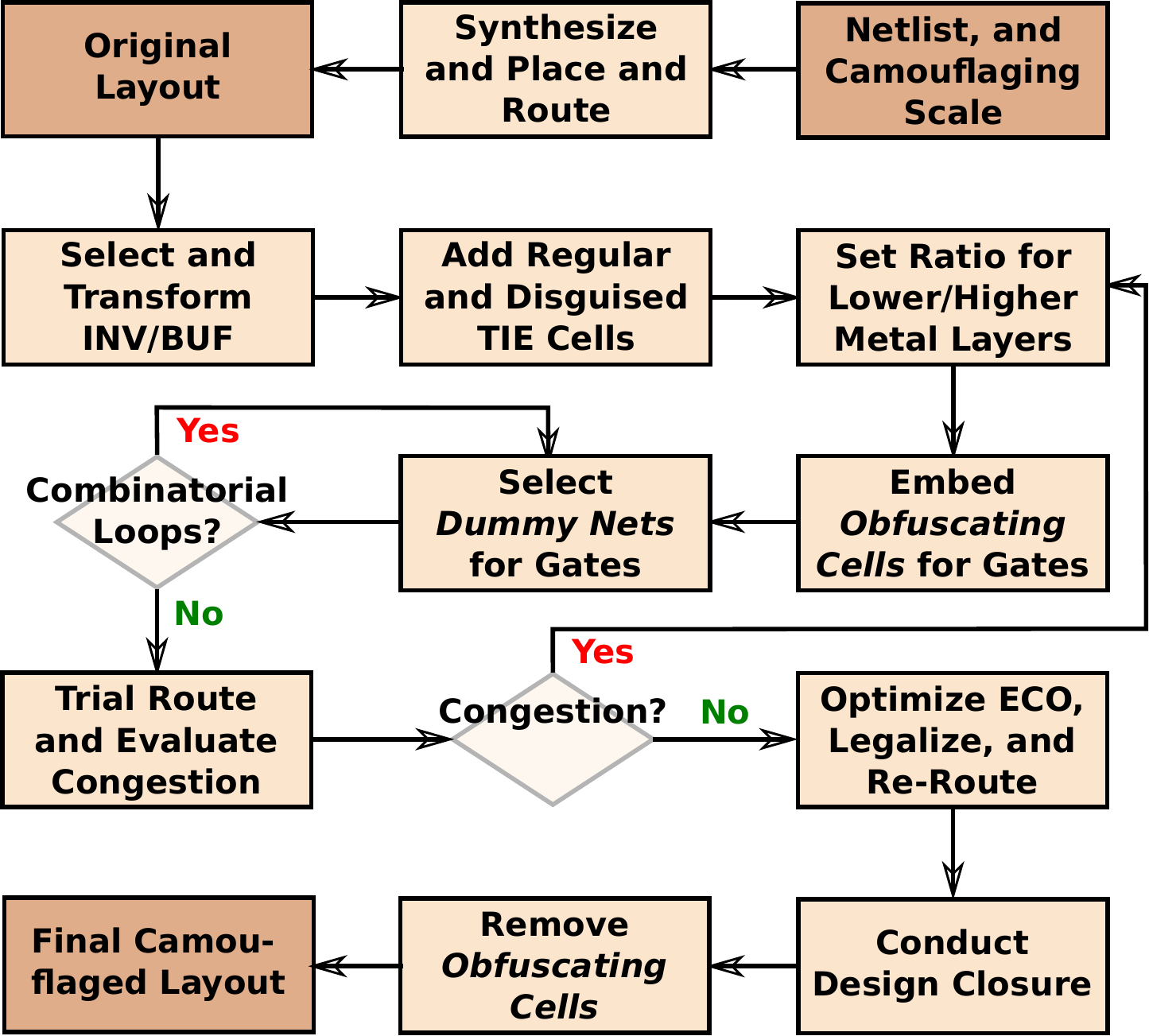}
\caption{Flow of our layout-level, BEOL-centric camouflaging methodology.}
\label{fig:flow}
\end{figure}

Here we provide an overview of our methodology (Fig.~\ref{fig:flow}), which can be easily
integrated into any design flow.
In this work, we implement our methodology for \textit{Cadence Innovus.}

Given a netlist,
      we initially synthesize, place, and route the design.
On this original layout,
we then apply our netlist transformations outlined above,
along with the insertion of TIE cells.
Recall that, for our scheme, we define \textit{camouflaging scale}, or \textit{LC scale} for short, as the ratio of obfuscated nets over all the nets.
Also recall that this is directly comparable to the ratio of camouflaged gates over all the gates, as we always obfuscate all the input
nets of any gate selected for camouflaging.
Next, depending on the camouflaging scale, we prepare for camouflaging in lower and higher metal layers, e.g., 30\% camouflaging in lower layers and 70\% camouflaging in
higher layers for an camouflaging scale of 100\% overall.  That so-called \textit{metal-layer ratio} for lower/higher layers and the camouflaging scale are provided by the user.
Then we randomly assign nets to be camouflaged at lower/higher layers, subject to these two parameters.

\begin{figure}[tb]
\centering
\subfloat[]{\includegraphics[height=2.75cm]{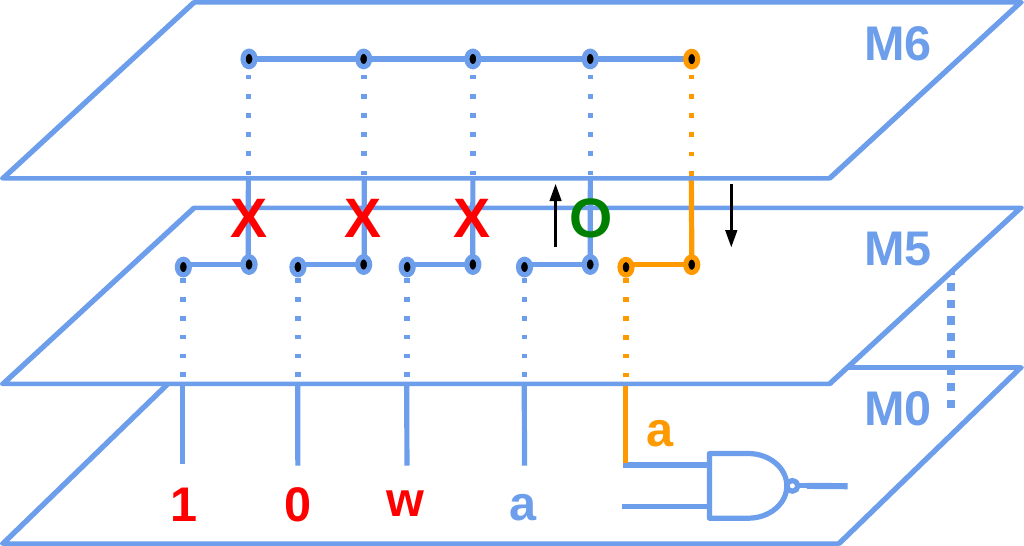}}\hfill
\subfloat[]{\includegraphics[height=2.8cm]{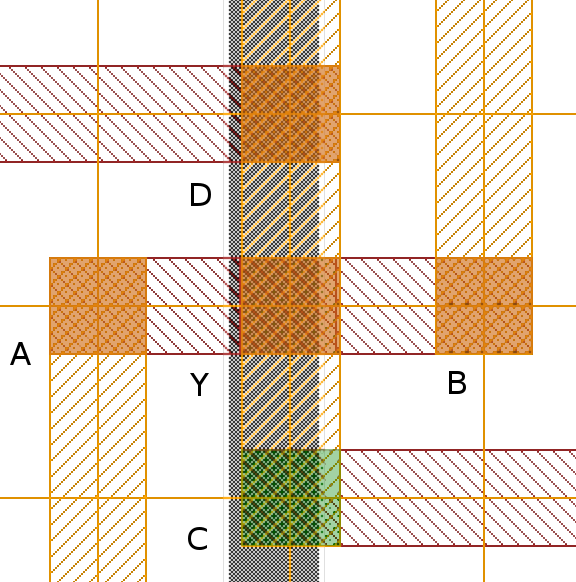}}
\caption{Wiring and vias for our \textit{obfuscating cell} implementing the BEOL-centric camouflaging in M5/M6.
Conceptional view in (a) and post-processed \textit{Innovus} layout snapshot in (b).
The dummy/real vias are indicated as red crosses/green circle in (a) and are colored in orange/green in (b).
In this example, the real input net is labeled as \textit{a} in (a).
In (b), the incoming wire for net \textit{a} connects to pin C, whereas the
outgoing wire connects to pin Y and from there to the input pin of the camouflaged gate (not illustrated).
The pins A, B, and Y reside in M6 (orange, vertical wiring), whereas pins C and D are set up in M5 (red, horizontal wiring).
Note that the actual wiring
depends on which vias are dummy and which are real; illustrated is one possible configuration.
}
\label{fig:final_primitive_layout}
\end{figure}

\begin{figure}[tb]
\centering
\includegraphics[width=.78\textwidth]{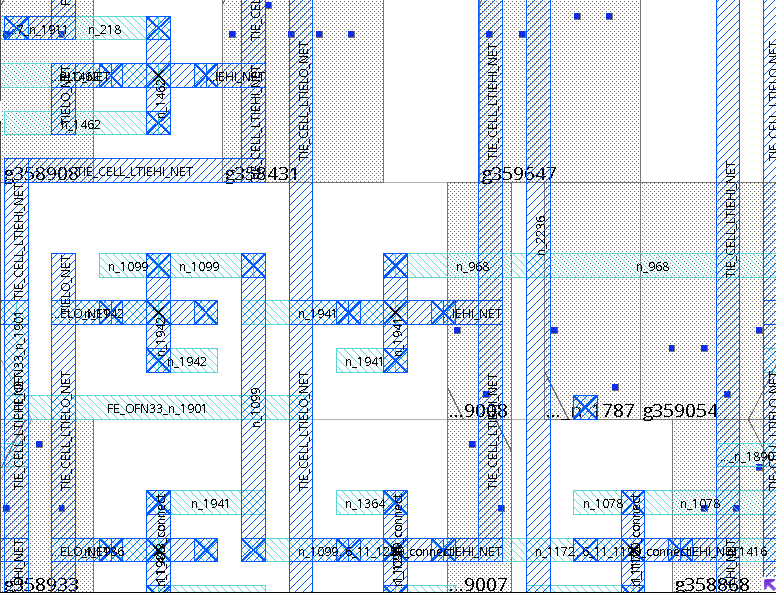}
\caption{\textit{Innovus} snapshot of multiple obfuscation cells, along with standard cells.
Colors are inverted, only M5/M6 layers are turned on,
and the outlines for obfuscation cells are turned off, all for visibility.}
\label{fig:layout_example}
\end{figure}

Next, we
insert and connect individual \textit{obfuscating cells} for all the inputs of each gate to be camouflaged---these
cells are essentially implementing our camouflaging primitives.
It is essential to understand that these customized obfuscating cells do not impact the FEOL layers---their sole purpose is to
enable the routing of all dummy and real wires
(Fig.~\ref{fig:final_primitive_layout}).
Hence, the physical design of these customized cells is tailored for routability, while their arrangement remains
flexible and unconfined concerning the already placed standard cells.
See Sec.~\ref{sec:obfuscating_cell} for more details on the physical design, and see
Fig.~\ref{fig:layout_example} for
a layout snapshot of multiple obfuscating cells in M5/M6.

As already indicated in Sec.~\ref{sec:concept},
we have to choose dummy nets carefully such that they are unique concerning each gate to be camouflaged.
We do so by applying a local spatial search around each gate's inputs; nearby nets/wires are preferably selected to limit the routing congestion.
We also check for combinatorial loops, which may have resulted during that process and
re-select dummy nets as required.

After embedding and connecting the obfuscating cells, we perform trial routing---now with all the dummy wires along with the real wires---and we
evaluate the congestion. 
In case of excessive congestion, the user is tasked to revise (i)~the metal-layer ratio and, if need be, (ii)~the die outline.
Regarding (i), we advocate to revise it toward more utilization of higher layers and less utilization of lower layers
for the following reasons.
First, we note that lower layers tend to become more utilized in any case, even when obfuscating cells are assigned to higher layers. That is
because the related dummy wires can also incur additional wiring throughout the lower layers.  Second, we note that obfuscating
cells and their vias inserted into the lower layers tend to require even more resources.
Regarding (ii), in case congestion remains after exploring various metal-layer ratios, we advocate to enlarge the die outline carefully
to gain more routing
resources.\footnote{An
alternative to gain routing resources is to employ additional metal layers, which may also be
more economical than enlarging die outlines~\cite{patnaik18_SM_ASPDAC}.}
      
Once we obtain an uncongested solution, we perform legalization and ECO optimization; the latter is 
also based on custom constraint rules, see
Sec.~\ref{sec:obfuscating_cell}.
At the same time, we also re-route the design.
Drivers are automatically up-scaled as needed to effectively/seemingly drive all the additional true/dummy wiring and the increased fanouts; see
also Sec.~\ref{sec:obfuscating_cell}.
We perform final design closure,
remove the obfuscating cells from the design,
re-extract the \textit{RC} data, stream out the GDSII data, and gather the final APD/PPA numbers.
Note that, after removal of the obfuscating cells,
all the true/dummy vias and related wiring remain as such in place of those cells.
Thus, the final APD/PPA numbers represent the nature of the camouflaged layout accurately.

\subsection{Physical Design of Obfuscating Cells}
\label{sec:obfuscating_cell}

We implement the obfuscating cells as customized cells, which only pertain to the BEOL.
To that end, and without loss of generality, we extend the LIB/LEF files of the
\textit{NanGate 45nm Open Cell Library}~\cite{nangate11}.
Next, we elaborate on the crucial aspects of the cells' physical design.
Besides the design described here for metal layers M5/M6, we
also implemented obfuscating cells
for M3/M4
in a similar manner.

\begin{figure}[tb]
\centering
\includegraphics[width=.4\textwidth]{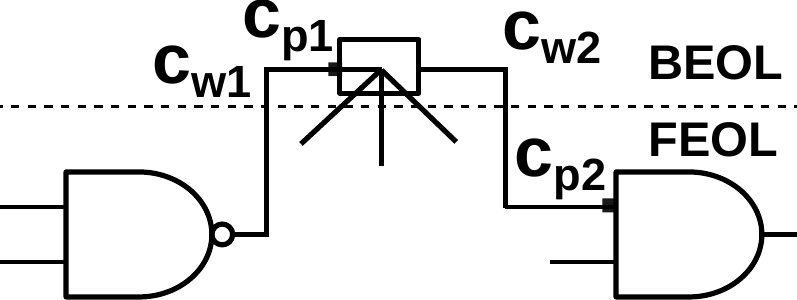}
\caption{The capacitance $c_{p1}$ for \textit{any} input pin of an obfuscating cell
is annotated such that it equals
   the wire capacitance $c_{w2}$ and the camouflaged gate's input-pin capacitance $c_{p2}$.}
\label{fig:obfus_pins_cap}
\end{figure}

\smallspaceenum
\begin{compactenum}

\item
The cells have four input and one output pin (Fig.~\ref{fig:final_primitive_layout}(b)).
These pins
have been set up in two metal layers: pins A, B, and Y reside
in M6,
whereas pins C and D are set up in M5.
We have chosen two different layers to
minimize the routing congestion---in exploratory experiments with all pins in M6, we observed overly high congestion, especially for camouflaging scales beyond 50\%.

\item
The dimensions of the pins ($0.14\times0.14\mu m$) and their offsets
are chosen such that the pins can be placed directly on the respective layer's tracks (thin yellow grid in
Fig.~\ref{fig:final_primitive_layout}(b)); this is to minimize routing congestion further.
Also, note the cell's grey
outline beneath the pins;
the pins
are partially located outside
the outline, which is supported.
The minimal width is solely to ease its visual differentiation from standard cells at design time.

\item
We define custom constraint rules
which prevent the pins of different obfuscating cells from overlapping during legalization.
These obfuscating cells
can, however, freely overlap with any standard cell without inducing routing conflicts. This is because standard cells have their pins exclusively in the lower metal layers.
Therefore, we can
camouflage the entire design without blocking any valuable standard-cell area.

\item
We leverage the timing and power characteristics of BUFX2, i.e., a buffer with a driving strength of 2.  
Note that a detailed library characterization 
is not required as obfuscating cells
implement BEOL wires and vias only.
However, since obfuscating cells are essentially interposed within the paths,
a \textit{capacitance annotation}
of subsequent loads
is crucial (Fig.~\ref{fig:obfus_pins_cap}).
Without such annotation,
the respective driver might become undersized,
as ECO optimization would only consider the wire routed from the driver toward
the obfuscating cell and the latter's BUFX2 input load, but not the camouflaged gate actually to be driven, which is ``masked'' by the obfuscating cell.
It is important to note that this annotation is to be applied for true nets as well as for dummy nets.
Otherwise, dummy nets might be identified as such by possibly undersized drivers.

\end{compactenum}

\section{Experimental Investigation}
\label{sec:experiments}

\subsection{Setup}
\label{sec:setup}

\textbf{Setup for physical design:}
We implement our methodology as custom \textit{TCL} scripts for
\textit{Cadence Innovus 16.15}.
Our procedures incur some runtime costs, as
presented in Table~\ref{tab:runtime_camouflaging_flow}.

\begin{table}[tb]
\centering
\scriptsize
\setlength{\tabcolsep}{1.2mm}
\caption{Runtime, in Seconds, for Our Camouflaging CAD Procedures}
\label{tab:runtime_camouflaging_flow}
\begin{tabular}{*{7}{c}}
\hline
\multirow{2}{*}{\textbf{Benchmark}} & 
\multirow{2}{*}{\textbf{Original}} & \textbf{20\%} 
& \textbf{40\%} & \textbf{60\%} 
& \textbf{80\%} & \textbf{100\%} \\ 
& 
& \textbf{LC Scale} 
& \textbf{LC Scale} & \textbf{LC Scale} 
& \textbf{LC Scale} & \textbf{LC Scale} \\ 
\hline 
\textit{aes\_core} & 498 
& 1,323 & 2,469 & 2,102 & 3,075 & 4,019  \\ %\hline
b14\_C & 178 
& 217 & 258 & 313 & 403 & 527  \\ %\hline
b15\_C & 256 
& 316 & 376 & 445 & 514 & 639  \\ %\hline
b17\_C & 603 
& 901 & 2,497 & 2,640 & 2,930 & 4,378  \\ %\hline
b22\_C & 428 
& 573 & 1,105 & 1,771 & 2,221 & 3,652 \\ %\hline
\textit{diffeq1} & 530 
& 819 & 1,576 & 2,733 & 4,349 & 6,256 \\ %\hline
\textit{square} & 451 
& 787 & 1,964 & 2,118 & 4,606 & 5,343 \\ \hline
\bf{Average} & 420.57 
& 705.14 & 1,463.57 & 1,731.71 & 2,385.43 & 3,544.86  \\ \hline
\end{tabular}
\\[1mm]
Benchmarks in italics are from the \textit{EPFL} suite~\cite{EPFL15}, others are from the \textit{ITC-99} suite.
\end{table}

We employ the public \textit{NanGate 45nm Open Cell
Library}~\cite{nangate11} with ten metal layers.
Synthesis is performed using \textit{Synopsys Design Compiler (DC)} for the slow process corner.
The APD analysis is carried out for
0.95V, 125$^\circ$C, and the slow process corner, along with a default input switching activity of 0.2---this
is a rather pessimistic setup, providing conservative results with ``safety margins'' and ``room for lowering cost.''
We obtain power and timing results using \textit{Innovus}.
We configure the initial utilization rates, i.e., for the original layouts, such that the routing congestion remains close to 0\%.

Recall that we implement the obfuscation cells as customized cells
in two versions, one for higher layers in M5/M6, and one for lower layers in M3/M4.
Unless stated otherwise, results are obtained with obfuscating cells embedded exclusively in M5/M6.
For experiments with both higher and lower layers being used,
we start with only lower layers being utilized, and revise that ratio,
without loss of generality and for the sake of simplicity, in steps of 1\% toward higher layers.

\textbf{Setup for security evaluation:}
We model ours as well as all camouflaging techniques proposed
in~\cite{rajendran13_camouflage, wang16_MUX, zhang16, nirmala16},
as outlined in~\cite{yu17}.
More specifically,
we model ours as individual 4-to-1 MUXes for all the inputs of each gate to be camouflaged, with all the related dummy/true wires connected to the MUXes' inputs and with two select/key
signals.
For the netlist transformation steps explained in Sec.~\ref{sec:fixed_values}, we select 50\% of all INVs/BUFs randomly and
transform them into various camouflaged gates randomly and
the remaining INVs/BUFs are camouflaged as is; both selections are also subject to the overall camouflaging scale.

For a fair evaluation, the same sets of gates are camouflaged across all camouflaging techniques:
for a given benchmark, gates are randomly selected once and then memorized.\footnote{Any
other technique such as \textit{maximum clique}~\cite{rajendran13_camouflage} could be applied here as well.
However, El Massad \textit{et al.}~\cite{el2019sat} observed
that maximum-clique selection, on average, cannot offer a better resilience than
random selection.}
Ten different sets are generated for each benchmark,
ranging from 10\% to 100\% camouflaging scale, in steps of 10\%.

For the scenario of malicious end-users,
we evaluate our scheme against powerful SAT-based attacks~\cite{subramanyan15,code_pramod,shen17,shen17_code} in Sec.~\ref{sec:security_evaluation_SAT}.
We note that these attacks were developed for
logic locking but are still applicable for our study---logic locking and camouflaging are closely related and can be transformed into one
another~\cite{yu17}.
We attribute the average runtime of the attacks as an empirical, yet essential, indicator for a design's resilience;
see also the discussion \textit{on the notion of practically secure camouflaging}
in Sec.~\ref{sec:security_evaluation_SAT}.

For the scenario of malicious fab employees, we evaluate our scheme against the powerful open-source proximity
attacks~\cite{wang2018cat,code_network_flow_attack} and~\cite{magana17,code_MAGANA_attack} in
Sec.~\ref{sec:security_evaluation_PA}.
We conduct experiments for layouts being split after M3 and M4, respectively, which
helps us to evaluate the significance of the split layer for our scheme.
For~\cite{wang2018cat,code_network_flow_attack}, we leverage the commonly applied metrics of \textit{correct connection rate (CCR)},
i.e., the ratio of correctly inferred BEOL connections over total BEOL connections, and 
\textit{Hamming distance (HD)} and \textit{output error rate (OER)}~\cite{patnaik18_SM_ASPDAC}; the latter both quantify the functional correctness of the inferred netlist.
HD and OER are computed using \textit{Synopsys VCS} using 100,000 patterns.
For~\cite{magana17,code_MAGANA_attack}, since it does
not provide any actual netlists, these metrics are not applicable; instead, we leverage the metrics defined in~\cite{magana17}, namely
\textit{virtual pins (vpins)}, \textit{E[LS]}, and \textit{figure of merit (FOM)}.\footnote{The term \textit{vpin} or
\textit{virtual pin} refers to the top-most, open, or ``dangling'' end of a FEOL wire, which is a hint for an adversary that a
via would be placed there to connect the FEOL with the undisclosed BEOL.  By definition,
\cite{magana17,code_MAGANA_attack} considers only \textit{vpins} related to two-pin nets.  \textit{E[LS]} is the number of
candidate \textit{vpins} to match with other \textit{vpins} over a specific region, i.e., the number of possible pairings
in the vicinity of some missing BEOL connection.  \textit{FOM} represents the number of possible pairings normalized over
the area of the specific region considered. Thus, when applied step wise across the FEOL layout and averaged,
\textit{FOM} serves to quantify the complexity of exploring all possible BEOL connections.  Since the
attack~\cite{magana17,code_MAGANA_attack} does not recover any actual netlists, these metrics are essentially indicators
for the complexity imposed on subsequent, yet-to-be-proposed attack steps to infer such netlists.}

All attacks are executed on a high-performance computing (HPC) facility.
Each computational node has two 14-core Intel
Broadwell processors (Xeon E5-2680), running at 2.4 GHz. Each node has 128 GB RAM in total, and 4 GB RAM are guaranteed (by the
\textit{Slurm} HPC scheduler) for each attack run. 
The CPU time-out (labeled as ``t-o'') is set to 48 hours.

\textbf{Benchmarks:} We conduct extensive experiments on traditional benchmarks suites (\textit{ISCAS-85}
and \textit{ITC-99}) and, for the first time, also on the
large-scale \textit{EPFL suite}~\cite{EPFL15} (using the original circuits, not the MIG versions).
Note that all benchmarks are combinatorial, but our approach can be directly applied to sequential designs as well.
We employ custom scripts to convert (i) the original netlists to \textit{bench} format with two-input gates, as required for the
SAT-based attack~\cite{code_pramod,subramanyan15},
and (ii) the post-layout netlists to \textit{rt} and \textit{out} format, as required for the proximity
attack~\cite{magana17,code_MAGANA_attack}.
We release the scripts for (ii) and other related materials in~\cite{webinterface2}.

\subsection{Layout Evaluation}
\label{sec:layout_evaluation}

\begin{figure}[tb]
\centering
\includegraphics[angle=-90,width=.95\textwidth]{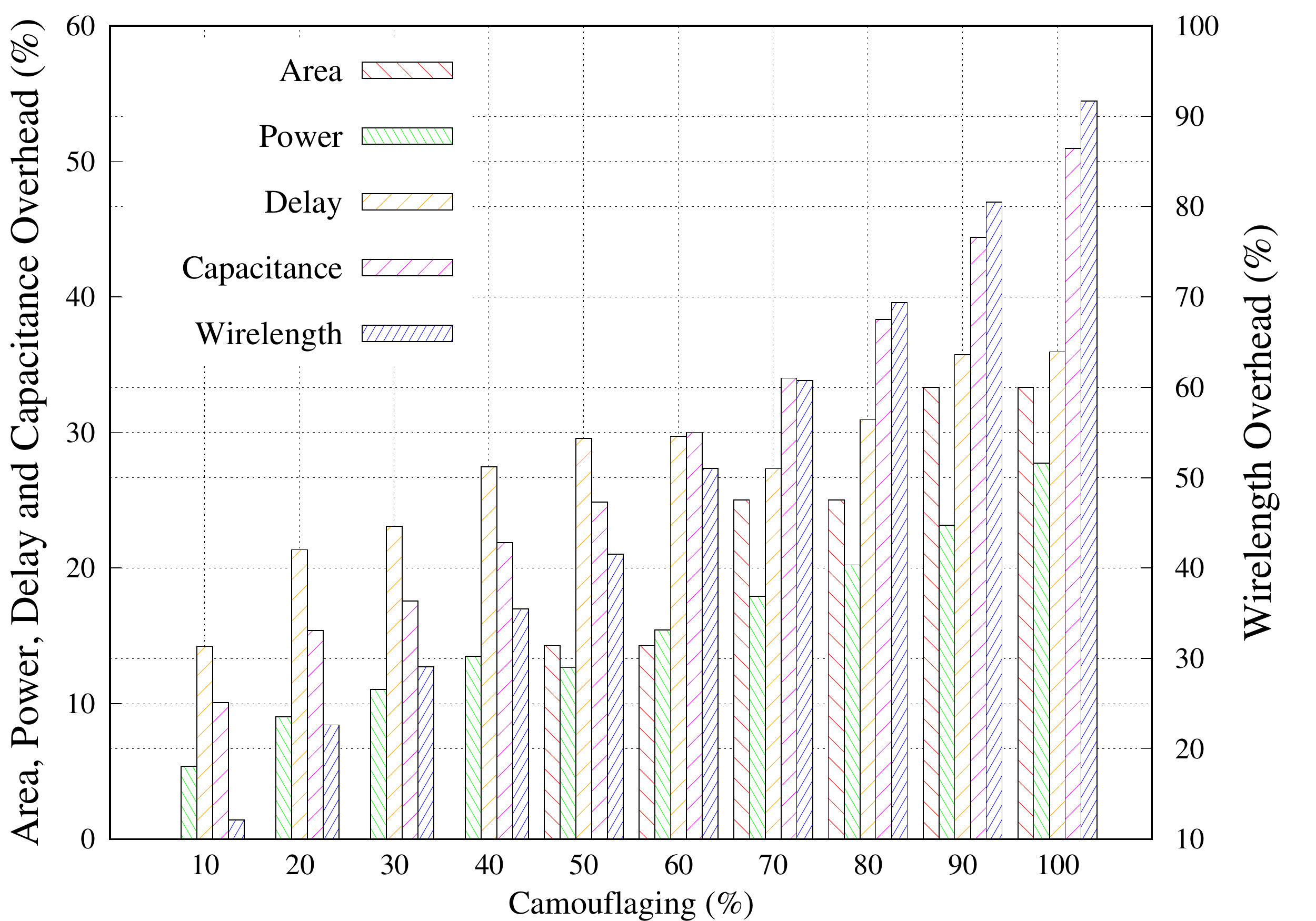}
\caption{Layout cost for \textit{EPFL} benchmark \textit{aes\_{core}}~\cite{EPFL15}.
The baseline is the original, unprotected layout.  The discrete and monotonous area cost is due to a step-wise up-scaling of die outlines as needed; see
text below.}
\label{fig:APD_aescore}
\end{figure}

We first report in detail on APD cost for our scheme, and
we then compare ours to various prior works.
Figure~\ref{fig:APD_aescore} illustrates our cost, along with wirelength
and load capacitance,
for the \textit{EPFL} benchmark \textit{aes\_{core}}~\cite{EPFL15}.
Table~\ref{tab:ppacomparison_timingaware} reports our cost for this and other selected benchmarks.
Note that we consider further benchmarks as well, for various other experiments, throughout this section.
Recall that results are in general obtained with obfuscating cells embedded only in M5/M6;
in Table~\ref{tab:APD_lower_metal_layers}, we also provide results for obfuscation cells being embedded in M5/M6 and M3/M4 at once.

\begin{table*}[tb]
\centering
\scriptsize
\setlength{\tabcolsep}{1.3mm}
\caption{Our Post-Layout, GDSII-Level Layout Cost}
\label{tab:ppacomparison_timingaware}
\begin{tabular}{*{23}{c}}
\hline

\multirow{2}{*}{\textbf{Benchmark}}
&
& \textbf{Utilization for}
&
& \multicolumn{3}{c}{\textbf{20\% LC Scale}}
&
& \multicolumn{3}{c}{\textbf{40\% LC Scale}}
&
& \multicolumn{3}{c}{\textbf{60\% LC Scale}}
&
& \multicolumn{3}{c}{\textbf{80\% LC Scale}}
&
& \multicolumn{3}{c}{\textbf{100\% LC Scale}}  \\
\cline{5-7}
\cline{9-11}
\cline{13-15}
\cline{17-19}
\cline{21-23}

&
& \textbf{Original Layout} 
&
& \textbf{Area} & \textbf{Power} & \textbf{Delay} 
&
& \textbf{Area} & \textbf{Power} & \textbf{Delay} 
&
& \textbf{Area} & \textbf{Power} & \textbf{Delay} 
&
& \textbf{Area} & \textbf{Power} & \textbf{Delay}  
&
& \textbf{Area} & \textbf{Power} & \textbf{Delay}  \\ \hline

\textit{aes\_{core}} &
& 0.4 &
& 0 & 9.02 & 21.35 
&
& 0 & 13.49 & 27.46 
&
& 14.29 & 15.44 & 29.72 
&
& 25 & 20.22 & 30.94 
&
& 33.33 & 27.76 & 35.72  \\ %\hline

b14\_C &
& 0.5 &
& 0 & 6.43 & 3.64 
&
& 0 & 9.88 & 5.01 
&
& 11.11 & 16.92 & 9.75
&
& 19.04 & 21.39 & 8.74 
&
& 19.04 & 38.46 & 14.76  \\ %\hline

b15\_C &
& 0.5 &
& 0 & 10.78 & 7.45 
&
& 0 & 11.19 & 10.14 
&
& 11.11 & 14.39 & 11.45 
&
& 25 & 17.27 & 15.74 
&
& 25 & 28.78 & 20.77  \\ %\hline

b17\_C &
& 0.5 &
& 11.11 & 7.16 & 7.19 
&
& 19.04 & 11.13 & 11.22 
&
& 25 & 16.93 & 14.64 
&
& 42.85 & 17.14 & 14.21 
&
& 42.85 & 28.29 & 16.92  \\ %\hline

b22\_C &
& 0.5 &
& 0 & 7.24 & 11.41 
&
& 0 & 13.11 & 14.37 
&
& 11.11 & 15.11 & 14.81 
&
& 19.04 & 25.27 & 15.11 
&
& 25 & 37.48 & 22.41  \\ %\hline

\textit{diffeq1} &
& 0.5 &
& 0 & 5.78 & 3.61 
&
& 0 & 10.64 & 6.54 
&
& 11.11 & 15.87 & 8.49 
&
& 11.11 & 24.19 & 16.79 
&
& 25 & 31.23 & 28.75  \\ %\hline

\textit{square} &
& 0.5 &
& 0 & 8.19 & 7.14 
&
& 11.11 & 9.98 & 13.06 
&
& 11.11 & 13.21 & 11.47 
&
& 19.04 & 18.3 & 16.42 
&
& 25 & 25.15 & 22.45  \\ \hline

\bf{Average} &
& 0.5 &
& 1.59 & 7.8 & 8.83 
&
& 4.31 & 11.35 & 12.54
&
& 13.55 & 15.41 & 14.33 
&
& 23.01 & 20.54 & 16.85 
&
& 27.89 & 31.02 & 23.11  \\ \hline
\end{tabular}
\\[1mm]
Layout cost are in percentage.
Obfuscating cells are embedded exclusively in M5/M6.  Benchmarks in
italics are from the \textit{EPFL} suite~\cite{EPFL15}, others are from the \textit{ITC-99} suite. All layouts are DRC-clean and devoid of
congestion.
\end{table*}

\textbf{On die area:} Recall that our obfuscating cells
impact only the metal layers; they
do not increase the standard-cell area.
Since wiring cannot be expressed by standard-cell area, the reported area cost is concerning \textit{die outlines} as explained next.
	
\begin{figure}[tb]
\centering
\subfloat[]{\includegraphics[width=.48\textwidth]{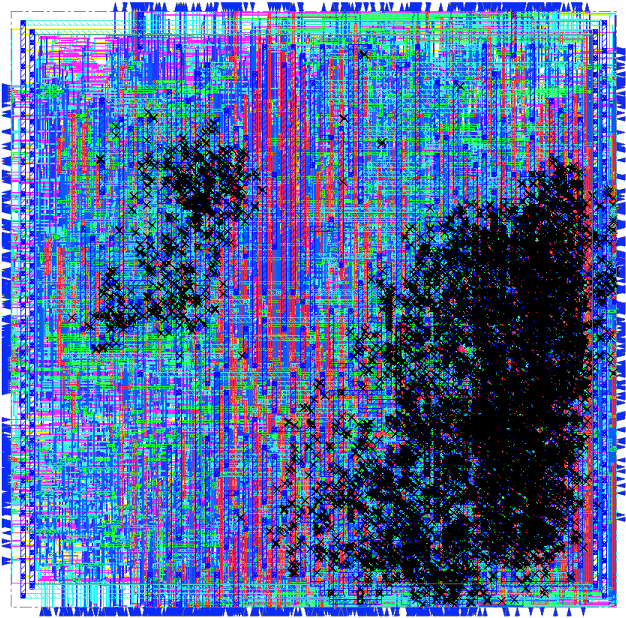}}\hfill
\subfloat[]{\includegraphics[width=.48\textwidth]{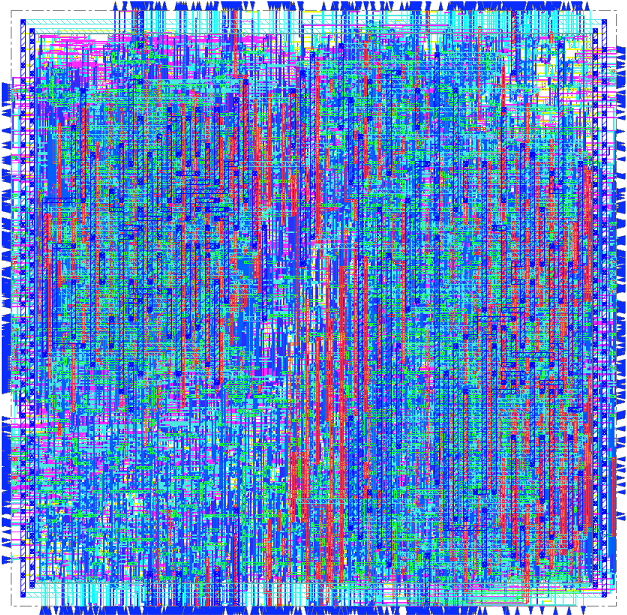}}\\
\caption{\textit{Innovus} snapshots for full-chip camouflaging of \textit{ITC-99 b15\_C}.
For visibility, colors are inverted.
(a) Camouflaging is implemented as 30\%/70\% in lower/higher layers, resulting in around 10,000 DRC issues (marked by crosses).
(b) For the same die outline, camouflaging is implemented as 20\%/80\% in lower/higher layers, resulting in a DRC-clean layout.
For (a), total over-congestion is reported as 10.4\%, with worst gcell over-congestion of 12.71\%, whereas for (b), these metrics are 
0.19\% and 0.43\%.}
\label{fig:b15_DRC_M4_M6_camo_100}
\end{figure}

Especially for large-scale camouflaging, with the additional wiring
imposed by our scheme
requiring significant routing resources, we can observe
routing congestion and DRC issues.
Recall that we first revise the metal-layer ratio, e.g., see Fig.~\ref{fig:b15_DRC_M4_M6_camo_100}, and, if need be,
scale-up die outlines.
In other words, we ultimately maintain that our large-scale camouflaged layouts are ``DRC-clean'' at the cost of larger outlines.
For simplicity, we scale-up the outlines by
decreasing the utilization rate in steps of
0.05 and 0.02, as appropriate.
For example, while relaxing the utilization from 0.5 to 0.45, the die area has to be increased by 11.11\%.

\begin{table}[tb]
\centering
\scriptsize
\setlength{\tabcolsep}{0.75mm}
\caption{Our Power (P), Delay (D) Cost for Camouflaging Only in M5/M6 (\textit{H}) Versus Camouflaging in M3/M4 and M5/M6 (\textit{L\&H})}
\label{tab:APD_lower_metal_layers}
\begin{tabular}{*{15}{c}}
\hline

\multirow{2}{*}{\textbf{LC Scale}}
&
& \multicolumn{2}{c}
{\textbf{b14\_C \textit{H}}} 
&
& \multicolumn{3}{c}
{\textbf{b14\_C \textit{L\&H}}} 
&
& \multicolumn{2}{c}
{\textbf{b15\_C \textit{H}}} 
&
& \multicolumn{3}{c}
{\textbf{b15\_C \textit{L\&H}}} 
\\
\cline{3-4}
\cline{6-8}
\cline{10-11}
\cline{13-15}

&
& \textbf{P} & \textbf{D} 
&
& \textbf{P} & \textbf{D} & \textbf{L/H}
&
& \textbf{P} & \textbf{D} 
&
& \textbf{P} & \textbf{D} & \textbf{L/H} 
\\
\hline

20\% & 
&
6.43 & 3.64 &
&
0.87 & -1.08 &
15/5 &
&
10.78 & 7.45 &  
&
2.94 & 1.46 &
15/5 
\\ %hline

40\% & 
&
9.88 & 5.01 & 
&
2.17 & 1.99 & 
10/30 &
&
11.19 & 10.14 &  
&
9.43 & 5.94 &
15/25
\\ %hline

60\% & 
&
16.92 & 9.75 & 
&
16.12 & 7.18 & 
20/40 &
&
14.39 & 11.45 &  
&
11.75 & 9.98 &
20/40
\\ %hline

80\% & 
&
21.39 & 8.74 & 
&
14.27 & 8.24 & 
30/50 &
&
17.27 & 15.74 &  
&
13.68 & 12.22 &
20/60
\\ %hline

100\% & 
&
38.46 & 14.76 & 
&
19.91 & 18.76 & 
30/70 &
&
28.78 & 20.77 &  
&
16.99 & 17.29 &
20/80
\\ \hline

\bf{Average} 
&
& 18.62 & 8.38
&
& 10.67 & 7.02 &
--
&
& 16.48 & 13.11 
&
& 10.96 & 9.38 &
--
\\ \hline
\end{tabular}
\\[1mm]
Layout cost and metal-layer ratios (L/H) are in percentage, with the latter being subject to the camouflaging scale (LC).
For a fair comparison, the same die outlines are considered for both the \textit{H} and \textit{L\&H} setups.
Note that the cost for the \textit{H} setup are the same as in Table~\ref{tab:ppacomparison_timingaware}.
\end{table}

The average area overheads in Table~\ref{tab:ppacomparison_timingaware} are not more than 14\% for up to 60\% camouflaging scale, whereas for
full-chip camouflaging, we note an average area cost close to 28\%.  Again, these
overheads enable DRC-clean layouts even for full-chip camouflaging; we believe that this is a justifiable cost.

Finally, while the transformation of 50\% of INV/BUF gates and the insertion of ``TIE cells in disguise`` both still incur some standard-cell area cost,
this cost is negligible compared to the die cost related to all the additional wiring. In fact, we have conducted additional experiments for full-chip camouflaging of the
	large-scale \textit{EPFL} benchmark \textit{aes\_core}~\cite{EPFL15} for
	ranges of 25--100\% of INVs/BUFs selected for transformation; we did not observe any additional die costs there.

\begin{figure*}[tb]
\centering
\subfloat[]{\includegraphics[width=.243\columnwidth]{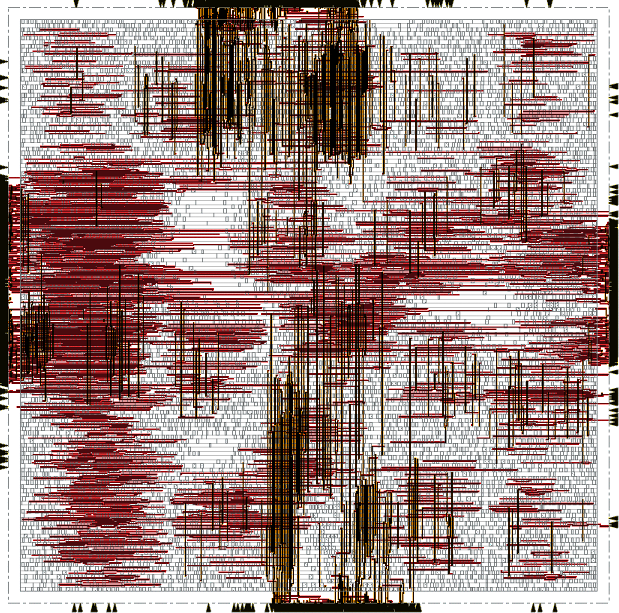}}\hfill
\subfloat[]{\includegraphics[width=.246\columnwidth]{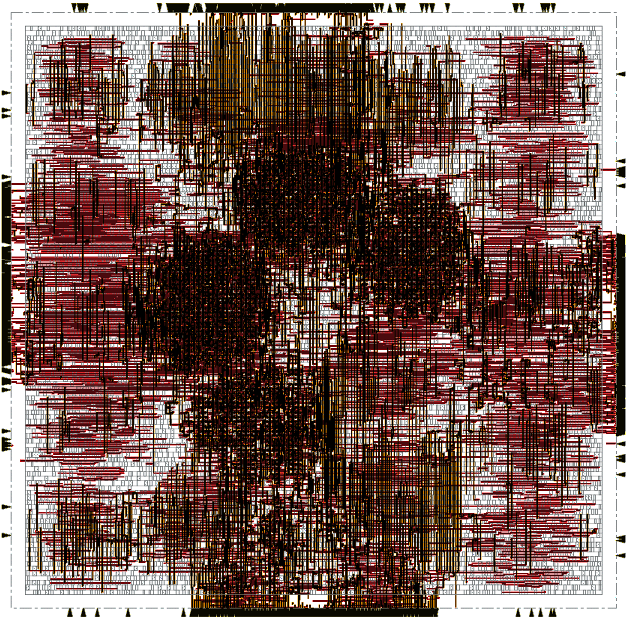}}\hfill
\subfloat[]{\includegraphics[width=.243\columnwidth]{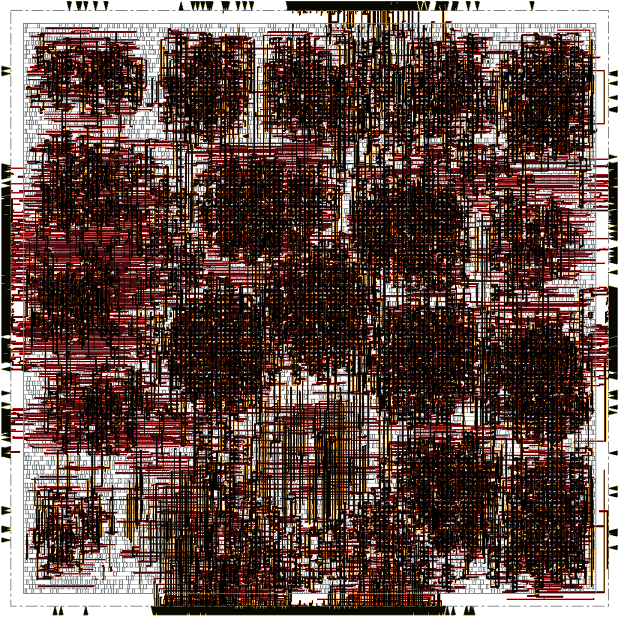}}\hfill
\subfloat[]{\includegraphics[width=.245\columnwidth]{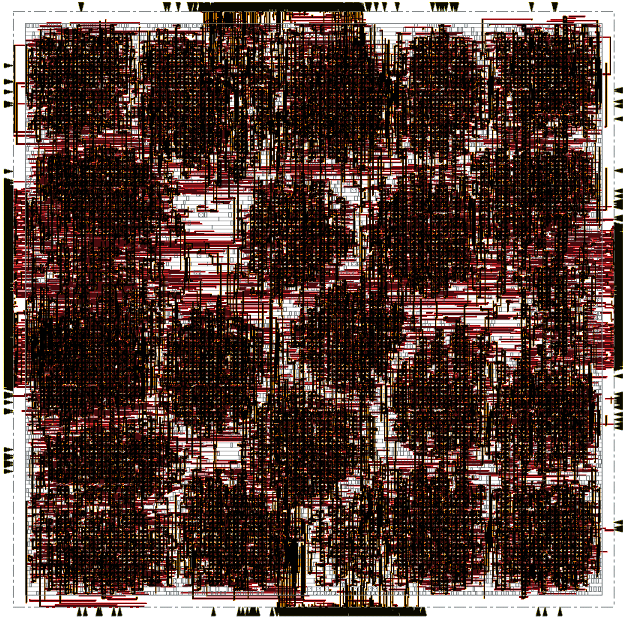}}\\
\caption{\textit{Innovus} snapshots of metal layers M5 and M6 for \textit{EPFL} benchmark \textit{aes\_{core}}~\cite{EPFL15} for (a) the original
	layout (i.e., 0\% LC scale),
(b) 20\% LC scale, (c) 60\% LC scale, and (d) 100\% LC scale.
Obfuscating cells are embedded exclusively in M5/M6.
Colors are inverted for visibility.
}
\label{fig:metal-AES}
\end{figure*}

\textbf{On power and performance:} As our camouflaging primitive incurs additional wiring, which becomes significant for large-scale camouflaging, an impact on
power and performance is expected.
See also Fig.~\ref{fig:metal-AES} for wiring snapshots across camouflaging ranges.

For camouflaging up until 60\% of the layout, the average power and delay overheads are moderate for most benchmarks, around
6--17\% for power and 4--15\% for delay (Table~\ref{tab:ppacomparison_timingaware}).
Only for \textit{EPFL} benchmark \textit{aes\_core}, delays are consistently larger, by approximately 15\%.
The overheads increase naturally toward full-chip camouflaging, but still follow linear growth trends.
We believe that this is due to the following:
\smallspaceenum
\begin{compactenum}
\item Besides the transformed INV/BUF gates, all gates remain as is.
Even for these transformed gates, cost are limited: for
experiments on varying ranges of INVs/BUFs transformed for full-chip camouflaging of the
large-scale \textit{EPFL} benchmark \textit{aes\_core}~\cite{EPFL15}, we observed variations for power and performance cost of less than 9\%.
That is, we experience no inherent overheads for the majority of gates and marginal overheads for transformed gates.
\item The lower resistance of higher metal layers, which are typically leveraged by most of the obfuscating-cell instances, helps to limit the delay cost.
\item The nets for fixed-values \textit{0} and \textit{1} are not switching and, thus, they neither increase power nor delay.
Also, while dummy nets are switching, they drive only wiring loads.
\item For large-scale camouflaging, the positive effects are eventually offset by a
steady increase of dummy nets and wiring for camouflaged gates,
thereby raising the capacitive loads and power consumption as well as routing congestion.
Since congestion can only be managed by re-routing in some detours, this lengthens parts of the wires further.
In turn, this also impacts power and delay.

\end{compactenum}
\smallspaceenum

Regarding the use of obfuscating cells in both lower and higher metal layers (Table~\ref{tab:APD_lower_metal_layers}), we observe the following. 
First, utilizing lower and higher layers at once (\textit{L\&H} setup) allows to reduce cost,
especially for power---in contrast to utilizing only the higher layers (\textit{H} setup),
we observe a reduction of 7.95\% and 5.52\% in power for the \textit{ITC-99} benchmarks b14\_C and b15\_C, respectively, and the
delay reduction is 1.36\% and 3.73\% for the same benchmarks.
To explain these trends, we also determine the load capacitance overheads for both setups:
for the \textit{H} setup, the capacitance overhead is about 62.78\%, whereas
this overhead is reduced to 29.99\% for the \textit{L\&H} setup.
Second, for large-scale camouflaging, the usability of lower layers becomes less; recall the related discussion in Sec.~\ref{sec:CAD_flow}.
Finally, recall that splitting at higher layers is considered beneficial~\cite{xiao15,patnaik18_SM_ASPDAC}, but here we cannot split above the
lowest layer containing some camouflaging primitives. Therefore, while the holistic use of lower and higher layers can help to reduce layout cost, it limits 
the practicality of split manufacturing. For a weaker threat model with the fab being \textit{trusted}, however, this approach would become more relevant.

\textbf{Comparison at RTL level and for small-scale camouflaging:}
Previous works typically report their cost only for small-scale camouflaging and based on RTL simulations, which seems impractical.
For example, Rajendran \textit{et al.}~\cite{rajendran13_camouflage} observe $\approx$60\%, $\approx$40\%, and $\approx$30\% increase in area,
power, and delay, respectively, when camouflaging 5\% of gates in the \textit{ISCAS-85} benchmark \textit{c7552}.
The approach of Wang \textit{et al.}~\cite{wang16_MUX} exhibits 50\% delay and 15\% area overhead already for 5\% camouflaging scale.
For another, more relevant example, Chen \textit{et al.}~\cite{chen18_interconnects} report
3.68\% delay and 3.93\% area overhead on average, i.e., when configuring their scheme for 3\% power budget/overhead
and while obfuscating only tens to few hundreds of wires
for the \textit{ITC-99} benchmarks.

\textbf{Comparison on physical-layout level:} Recall that our work is one of the very few to evaluate camouflaging on placed-and-routed GDSII layouts.
When contrasting to a previous study, conducted by Malik \textit{et al.}~\cite{malik15}, we notice significantly higher overheads for them.
We note that Malik \textit{et al.\ }implement and evaluate their approach
exclusively for one \textit{AES S-box},
which has a far lower number of gates, namely 421, when compared to all the benchmarks we consider.
However, we argue that a qualitative comparison is still fair as Malik \textit{et al.\ }use the same \textit{NanGate} library~\cite{nangate11}.
They reported
overheads of 7.09$\times$, 6.45$\times$, and 3.12$\times$ for area, power, and delay (APD), respectively, for their case study.
Moreover, Malik \textit{et al.\ }indicate themselves that cost will increase for larger circuits.

Besides a provably secure scheme, Li \textit{et al.}~\cite{li16_camouflaging}
propose two ``regular'' camouflaging primitives (based on obfuscated contacts), called \textit{STF-type} and \textit{XOR-type}.
As Li \textit{et al.\ }report only gate-level cost for their primitives, we conduct a layout-level comparison for full-chip camouflaging ourselves (Table~\ref{tab:ppacomparison_new}).
Here we first constrain synthesis accordingly (recall Sec.~\ref{sec:LC_FEOL}), and then we map their gate-level cost to all the
gates---this extrapolation is
conservative as it implies only a linear scaling even for full-chip camouflaging.
While the STF-type primitive becomes a contender to ours in terms of power and
delays, it still incurs 23.88\% larger die-area cost on average.

For Zhang's work~\cite{zhang16}, as its MUX-based primitive is not made available, we implement it ourselves, by using MUXes and INVs
for the primitives, and primary inputs for their key bits, and we perform a detailed layout-level evaluation.
For example, we observe 464\%, 638\%, and 63\% increase for APD, respectively, when camouflaging all
the gates for the \textit{ISCAS-85} benchmark \textit{c7552}.
These numbers are 13.92$\times$, 37.51$\times$, and 4.47$\times$ higher than ours for the same scenario.

On average, across all \textit{ISCAS-85}, \textit{ITC-99}, and \textit{EPFL} benchmarks,
our scheme incurs only 32.55\%, 24.96\%, and 19.06\% APD cost for full-chip camouflaging.

\begin{table}[tb]
\centering
\scriptsize
\setlength{\tabcolsep}{0.4mm}
\caption{Area (A), Power (P), and Delay (D) for Full-Chip Camouflaging}
\label{tab:ppacomparison_new}
\begin{tabular}{*{16}{c}}
\hline

\textbf{Bench-}
& \multicolumn{3}{c}{\textbf{XOR-type~\cite{li16_camouflaging}}} 
&
& \multicolumn{3}{c}{\textbf{STF-type~\cite{li16_camouflaging}}} 
&
& \multicolumn{3}{c}{\textbf{Threshold~\cite{erbagci16}}}
&
& \multicolumn{3}{c}{\textbf{Ours}}
\\
\cline{2-4}
\cline{6-8}
\cline{10-12}
\cline{14-16}

\textbf{mark}
& \textbf{A} & \textbf{P} & \textbf{D} 
&
& \textbf{A} & \textbf{P} & \textbf{D} 
&
& \textbf{A} & \textbf{P} & \textbf{D}  
&
& \textbf{A} & \textbf{P} & \textbf{D}   
\\ 
\hline

c432 
& 66.48 & 11.43 & 42.2  
&
& 54.31 & 20.37 & 6.4 
&
& 140 & 8 & 96 
&
& 27.27 & 25.46 & 15.41   
\\ %hline

c5315 
& 92.33 & 33.81 & 50.56 
&
& 62.18 & 38.31 & 7.29 
&
& 200  & 10 & 76
&
& 50 & 10.19 & 9.94   
\\ %hline

c7552 
& 80.86 & 58.94 & 29.64  
&
& 62.53 & 63.42 & 14.86  
&
& 175 & 9 & 90 
&
& 33.33 & 17.01 & 14.09   
\\ %hline

b14\_C
& 40.43 & -5.41 & 41.6 
&
& 42.23 & 13.11 & 22.9 
&
& N/A & N/A & N/A 
&
& 19.04 & 38.46 & 14.76   
\\ %hline

b15\_C 
& 67.07 & 16.36 & 37.13 
&
& 65.4 & 30.77 & 19.54 
&
& N/A & N/A & N/A 
&
& 25 & 28.78 & 20.77   
\\ %hline

b17\_C 
& 71.40 & 25.12 & 55.97 
&
& 59.91 & 11.99 & 14.39
&
& N/A & N/A & N/A  
&
& 42.85 & 28.29 & 16.92   
\\ %hline

b22\_C
& 46.37 & 7.67 & 47.65
&
& 43.09 & 25.17 & 25.87
&
& N/A  & N/A  & N/A 
&
& 25 & 37.48 & 22.41   
\\ \hline

\bf{Average} 
& 66.42 & 21.13 & 43.54 
&
& 55.66 & 29.02 & 15.89
&
& 171.67 & 9 & 87.33
&
& 31.78 & 26.52 & 16.33
\\ \hline
\end{tabular}
\\[1mm]
We consider selected \textit{ISCAS-85} and \textit{ITC-99} benchmarks, as done in the listed prior art.
All layout
cost are in percentage.
We evaluate cost for \cite{li16_camouflaging} ourselves, while constraining synthesis and applying a linear scaling of 
baseline camouflaging primitives cost, reported in \cite{li16_camouflaging}, capturing all gates.
Cost for \cite{erbagci16} are quoted.
N/A means not available, i.e., the respective authors did not consider that benchmark.
\end{table}

\textbf{Comparison with threshold-dependent camouflaging:} Nirmala \textit{et al.}~\cite{nirmala16} proposed a promising concept of
threshold-dependent camouflaging
switches.
Since their primitive and related libraries are not made available, we apply our own layout-level evaluation as follows.
First, we conduct a constrained synthesis run (recall Sec.~\ref{sec:LC_FEOL}) using the \textit{NanGate} library~\cite{nangate11}.
Then, based on the overhead numbers reported for their primitive, we scale up the cost accordingly for all the camouflaged gate instances.
Again, while this is an extrapolating approach, we argue that it is fair, as we reflect truthfully the different overheads for
different types of gates camouflaged.
   We find that the approach by Nirmala \textit{et al.} will incur significant APD cost for full-chip camouflaging: $\approx$1,360\%, $\approx$1,266\%, and $\approx$100\%,
respectively.
For the work of Collantes \textit{et al.}~\cite{collantes16}, we observe a linear trend in the reported layout cost;
simply extrapolating those
numbers for full-chip camouflaging would translate to $\approx$78\% and $\approx$147\% for power and delay, respectively. 
Finally, for the work of Erbagci \textit{et al.}~\cite{erbagci16}, we quote their numbers in Table~\ref{tab:ppacomparison_new}.

\textbf{Comparison with provably secure camouflaging:} Recall that schemes like~\cite{xie16_SAT,yasin16_CamoPerturb,li16_camouflaging} rely on additional circuitry to
protect individual, selected gates, wires, or outputs. Such circuitry can incur a high cost, especially for area and power.
Using the benchmarks provided by \textit{Amir et al.}~\cite{amir2018development} on \textit{trust-hub.org},
we conduct a layout-level analysis for \textit{Anti-SAT}~\cite{xie16_SAT},
    namely for \textit{Anti-SAT} with random logic locking (RLL) and for
\textit{Anti-SAT} with secure logic locking (SLL)~\cite{yasin16_locking_TCAD}.\footnote{Recall
that locking and camouflaging are interchangeable, especially w.r.t. SAT-based attacks~\cite{yu17}, rendering this analysis relevant for our work.}
	We perform a regular design flow here,
	using the \textit{Nangate} 45nm library~\cite{nangate11} at the slow corner and with timing constrained to 4ns.
When comparing these schemes to our full-chip camouflaging
(Table~\ref{tab:APD_comparison_with_Anti_SAT}),
we observe that average overheads are notably larger, namely by
8.95$\times$ and 9.59$\times$ for area
and by
23.66$\times$ and 26.05$\times$ for power, respectively;
only delay overheads are less.
We note that these cost arise for protecting single outputs and increase further if we would protect
more outputs.

In any case, we emphasize again that our notion of full-chip camouflaging is different from such schemes (including functional
		obfuscation~\cite{bypass-attack2017} discussed next),
see also \textit{on the notion of practically secure camouflaging} further below.

\begin{table}[tb]
\centering
\scriptsize
\setlength{\tabcolsep}{0.8mm}
\caption{Comparison of Layout-Level Cost
with \textit{Anti-SAT}~\cite{xie16_SAT,yasin16_locking_TCAD,amir2018development} on selected \textit{ISCAS-85} benchmarks} 
\label{tab:APD_comparison_with_Anti_SAT}
\begin{tabular}{*{13}{c}}
\hline
\multirow{2}{*}{\textbf{Benchmark}}
&
& \multicolumn{3}{c}
{\textbf{Anti-SAT+RLL}}%~\cite{xie16_SAT,amir2018development}}} 
&
& \multicolumn{3}{c}
{\textbf{Anti-SAT+SLL}}%~\cite{xie16_SAT,amir2018development}}} 
&
& \multicolumn{3}{c}
{\textbf{Ours}}
\\
\cline{3-5}
\cline{7-9}
\cline{11-13}

&
& \textbf{A} & \textbf{P} & \textbf{D} 
&
& \textbf{A} & \textbf{P} & \textbf{D} 
&
& \textbf{A} & \textbf{P} & \textbf{D}   
\\
\hline

c880 & 
&
588.48 & 678.13 & 5.29 &
&
568.42 & 618.15 & 7.89 &
&
27.27 & 23.28 & 19.63   
\\ %hline

c1355 & 
&
480.15 & 410.09 & 16.45 &
&
500.44 & 418.83 & 16.13 &
&
40 & 9.48 & 11.62   
\\ %hline

c1908 &
&
479.06 & 526.91 & 2.74 &
&
552.91 & 644.49 & 3.82 &
&
40 & 21.11 & 16.46   
\\ %hline

c3540 &
&
193.87 & 162.05 & 0.92 &
&
290.07 & 335.75 & 0.31 &
&
50 & 10.34 & 10.24    
\\ %hline

c5315 &
&
230.15 & 212.91 & 2.08 &
&
224.59 & 193.83 & 0.65 &
&
50 & 10.19 & 9.94    
\\ %hline

c7552 &
&
182.22 & 173.21 & -4.42 &
&
170.92 & 170.64 & -4.83 &
&
33.33 & 17.01 & 14.09   
\\ \hline

\textbf{Average} & 
&
358.99 & 360.55 & 3.84 &
&
384.56 & 396.95 & 3.99 &
&
40.1 & 15.24 & 13.66  
\\ \hline
\end{tabular}
\\[1mm]
All layout cost are in percentage.
We use \textit{trust-hub.org} benchmarks~\cite{amir2018development} as follows.
For \textbf{Anti-SAT+RLL}, we use c880-NR4360, c1355-NR3790, c1908-NR3550, c3540-NR4060, c5315-NR7900, and c7552-NR8770.
For \textbf{Anti-SAT+SLL}, we use c880-NS4360, c1355-NS3790, c1908-NS3550, c3540-NS4060, c5315-NS7900, and c7552-NS8770.
All protected designs are configured for iso-performance at 4ns.
\end{table}

\textbf{Comparison with functional obfuscation:}
Xu \textit{et al.}~\cite{bypass-attack2017} proposed binary decision diagram (BDD)-based obfuscation
which operates at the functional level, in contrast to other approaches working at the netlist level.
It was shown that their scheme can be tailored to remain resilient against SAT-based attacks as well ``removal attacks''~\cite{bypass-attack2017,yasin17_TETC}.
As above, we evaluate two approaches of \cite{bypass-attack2017} at layout level, namely random obfuscation and \textit{Anti-SAT}-inspired
obfuscation~\cite{amir2018development}.
Again, expect for delay, we observe that average overheads are significantly larger, namely by
92.39$\times$ and 76.57$\times$ for area
and
197.1$\times$ and 246.7$\times$ for power, respectively
(Table~\ref{tab:APD_comparison_BDD_obfuscation}).

\begin{table}[tb]
\centering
\scriptsize
\setlength{\tabcolsep}{0.6mm}
\caption{Comparison of Layout-Level Cost with \textit{BDD-Based} Obfuscation~\cite{bypass-attack2017,amir2018development} on Selected \textit{ISCAS-85} Benchmarks}
\label{tab:APD_comparison_BDD_obfuscation}
\begin{tabular}{*{13}{c}}
\hline

\multirow{2}{*}{\textbf{Benchmark}}
&
& \multicolumn{3}{c}
{\textbf{Random Obfus.}}%~\cite{bypass-attack2017,amir2018development}}}
&
& \multicolumn{3}{c}
{\textbf{Anti-SAT Obfus.}}%~\cite{bypass-attack2017,amir2018development}}}
&
& \multicolumn{3}{c}
{\textbf{Ours}}
\\
\cline{3-5}
\cline{7-9}
\cline{11-13}

&
& \textbf{A} & \textbf{P} & \textbf{D} 
&
& \textbf{A} & \textbf{P} & \textbf{D} 
&
& \textbf{A} & \textbf{P} & \textbf{D}   
\\
\hline

c880 & 
&
5,166.86 & 6,351.69 & 8.14 &
&
5,289.19 & 6,536.58 & 9.44 &  
&
27.27 & 23.28 & 19.63
\\ %hline

c1908 & 
&
5,727.6 & 5,973.24 & 1.96 &
&
7,078.33 & 7,349.63 & 2.74 &  
&
40 & 21.11 & 16.46 
\\ %hline

c3540 & 
&
2,803.79 & 2,111.32 & 1.76 &
&
2,807.78 & 2,119.05 & 2.32 &
&
50 & 10.34 & 10.24    
\\ %hline

c5315 & 
&
76.89 & 45.49 & 7.26 &
&
51.63 & 39.59 & 6.67 &
&
50 & 10.19 & 9.94   
\\ %hline

c7552 & 
&
109.30 & 63.49 & 0.81 &
&
133.57 & 97.04 & -0.33 &
&
33.33 & 17.01 & 14.09 
\\ \hline

\textbf{Average} & 
&
3,706.84 & 4,040.95 & 4.72 &
&
3,072.1 & 3,228.38 & 4.17 &
&
40.12 & 16.38 & 14.07   
\\ \hline
\end{tabular}
\\[1mm]
All layout cost are in percentage.
We use \textit{trust-hub.org} benchmarks~\cite{amir2018development} as follows.
For \textbf{Random Obfus.} we use c880-BR320, c1908-BR320, c3540-BR320, c5315-BR320, and c7552-BR320. 
For \textbf{Anti-SAT Obfus.} we use %c432-BS320,
c880-BS320, c1908-BS320, c3540-BS320, c5315-BS320, and c7552-BS320.
Note that \textit{trust-hub.org}~\cite{amir2018development} does not provide the related c1355 circuits.
All protected designs are configured for iso-performance at 4ns.
\end{table}

\subsection{Security Evaluation: Malicious End-Users, SAT-Based Attacks} \label{sec:security_evaluation_SAT}

\textbf{On the notion of practically secure camouflaging:}
Recall that the primary objective
for this work is large-scale camouflaging, and
an important observation is that this achieves \textit{practically secure camouflaging}.
That is, by camouflaging up to 100\% of the layout,
we can induce significant computational efforts for SAT-based attacks,
without leveraging additional provably secure structures.

Still, it is not straightforward to prove beforehand to what extent large-scale camouflaging will
render a layout
resilient without actually leveraging a
SAT solver's capabilities for de-camouflaging attacks.
Li \textit{et al.}~\cite{li16_camouflaging} have shown that de-camouflaging efforts scale \textit{on average} with (i)~the solution space $C$ concerning all possible
functionalities of the whole design and (ii)~the Hamming distances among those different functionalities.
A further theoretical evaluation of large-scale camouflaging is difficult since $C$
depends on (i)~the number and composition of possible functionalities supported by the camouflaging primitives, (ii)~the number of camouflaged gates,
(iii) the selection of camouflaged gates, and (iv)~the interconnectivity of the design, all at the same time.
For example, designs containing XOR/XNOR and/or multipliers are harder to de-camouflage in practice than other designs~\cite{yu17,subramanyan15}.

\begin{figure}[tb]
\centering
\includegraphics[width=.80\columnwidth]{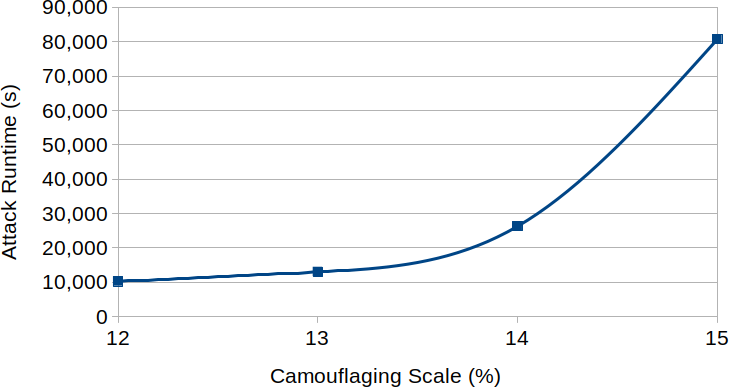}
\caption{Average runtimes of
\cite{subramanyan15,code_pramod} for
	\textit{aes\_core}, sampled over 70 different sets of randomly gates being camouflaged across the plotted range.
\label{fig:aes_scaled}
}
\end{figure}

In short, while one can easily estimate the upper bound of $C$,
   this may not reflect on the actual efforts required for successful attacks.
Hence, we next resort to an empirical but comprehensive study.
As we observe in this study,
   we can indeed expect prohibitive runtimes for large-scale camouflaging.
More specifically, e.g.,
we observe a polynomial trend for attacking the benchmark \textit{aes\_core}, setting in around 14\% camouflaging
(Fig.~\ref{fig:aes_scaled}).
Similar observations have also been
made by Yu \textit{et al.}~\cite{yu17}, albeit for a different primitive and the much smaller benchmark \textit{c432} with 209 two-input gates; even
such a small layout could not be resolved within three days.

\textbf{Comparison with prior art on large-scale camouflaging:}
In Table~\ref{tab:SAT_pramod}, we list runtimes for the seminal SAT-based attack~\cite{subramanyan15,code_pramod},
evaluating our camouflaging primitive and those
of~\cite{rajendran13_camouflage, wang16_MUX, zhang16, nirmala16}.
Note that these prior studies did not report on any SAT-based attacks for large-scale camouflaging themselves.
Also, recall that we camouflage the same sets of gates across all techniques for a fair comparison for each run, and that we report average
runtimes over ten such runs.

For our scheme, we do not observe any significant differences for the average SAT-based attack runtimes when considering
different ranges of INVs/BUFs being transformed (Sec.~\ref{sec:fixed_values}). This
is because, for cases with less INVs/BUFs being selected for transformation, more other gates are selected for camouflaging, while the
overall count of camouflaged gates remains the same; recall that we use the
camouflaging scale as design-wide ``knob'' for all of our experiments.

We note that none of the layouts can be de-camouflaged within 48 hours once full-chip camouflaging is applied. In fact, almost all layouts remain already
resilient beyond 40\% camouflaging scale; the only exception arises for the \textit{ISCAS-85} benchmark \textit{c7552} when
the most simple camouflaging primitive~\cite{zhang16} is leveraged.
We ran further exploratory attacks for 7 days on large-scale camouflaging using our primitive---without observing any improvement for the attack.
For a more meaningful comparison, we also consider relatively small camouflaging scales, i.e., 10\% up to 30\%.
Here, the primitive by Zhang \textit{et al.}~\cite{zhang16} appears as the weakest and that by Wang \textit{et al.}~\cite{wang16_MUX} as the most
resilient.
Our primitive is next only to those of~\cite{wang16_MUX,nirmala16}.
However, it is
also important to recall that ours incurs significantly less APD cost (Sec.~\ref{sec:layout_evaluation}).

We also leverage the \textit{Double DIP} attack, provided by Shen \textit{et al.}~\cite{shen17,shen17_code}.
The crux of their attack is that it can rule out at least
two incorrect keys during the application of one single \textit{distinguishing input pattern (DIP)} for the oracle.
While conducting the same set of experiments as above, we observe that the average runtimes are even higher across all benchmarks and camouflaging schemes (Table~\ref{tab:SAT-double-DIP}).
For example, for 
the \textit{ITC-99} benchmark \textit{b14\_C} with
10\% camouflaging applied using our primitive, we observe
$\approx$137 seconds for~\cite{code_pramod}, but $\approx$402 seconds for~\cite{shen17_code}. 
Thus, we conclude that \textit{Double DIP}, while successful for attacking dedicated protection schemes
such as \textit{SARLock}
cannot cope well with large-scale camouflaging schemes in general.

Overall, we are \textit{not} claiming that large-scale camouflaging, whether based on prior art or our primitive, cannot be resolved \textit{eventually} using SAT-based
attacks.
Rather, we provide strong empirical evidence that practically secure camouflaging, i.e., once 50--100\% of all gates are camouflaged, imposes prohibitive
computational cost on SAT solvers.

\textbf{On the (lack of) resilience for~\cite{chen18_interconnects}:}
Since our work is inspired to some degree by that of Chen \textit{et al.}~\cite{chen18_interconnects}, it seems imperative to investigate the
resilience of their scheme as well, especially since Chen \textit{et al.} did not consider any SAT-based attacks themselves.
In Table~\ref{tab:chen18_attack}, we note that the SAT-based attack \cite{subramanyan15, code_pramod}
can distinguish between real and dummy interconnects within relatively short runtimes, no matter how many
interconnects are obfuscated, i.e., at least concerning the delay constraints defined in~\cite{chen18_interconnects}.
That is, an overly constrained
obfuscation as proposed in~\cite{chen18_interconnects} cannot withstand powerful SAT-based attacks.
Similar observations have been also reported by Yu \textit{et al.}~\cite{yu17}.

\subsection{Security Evaluation: Fab Adversaries, Proximity Attacks}
\label{sec:security_evaluation_PA}

Similar to the observations made
for SAT-based
attacks, here we find that our scheme of obfuscating the interconnects---along with all the additional wiring going through both the FEOL and 
BEOL---serves also well to hinder proximity attacks. That is essentially due to vast numbers of wires being cut at the split layer, creating a
large solution space for the attacks to work on.
Besides, recall that the metrics used in this section have been introduced in Sec~\ref{sec:setup}.

We execute the seminal network-flow attack~\cite{wang2018cat,code_network_flow_attack} on different layouts for major camouflaging scales and for two split layers, M3 and M4, on selected
\textit{ISCAS-85} benchmarks (Table~\ref{tab:CCR_results_NFA}).\footnote{We execute the attack also for the larger
\textit{ITC-99} benchmarks, but it failed to conclude within 48 hours.}
As indicated, we observe that our scheme is effective in mitigating this attack.
The average CCR observed for original layouts is significantly reduced for fully camouflaged layouts, namely by 7.48$\times$ and 2.41$\times$ when split after M3
and M4, respectively.
This significant reduction in CCR is corroborated by the increase of cut wires to handle when attacking our scheme, reported as
cut inputs (CI)~\cite{code_network_flow_attack} in Table~\ref{tab:CCR_results_NFA}.
For example for the \textit{ISCAS-85} benchmarks
\textit{c5315} and \textit{c7552}, CI is increased by 4.61$\times$ and 7.98$\times$, respectively, when comparing fully camouflaged to regular
layouts, both split after M3.
Again, this increase in CI is due to the fact our scheme routes all nets related to the camouflaging primitive through higher layers.

\begin{table*}[tb]
\centering
\scriptsize
\setlength{\tabcolsep}{1.95mm}
\caption{Average Runtimes (in Seconds) for the Seminal SAT-Based Attack~\cite{subramanyan15,code_pramod}
on Selected \textit{ISCAS-85} and \textit{ITC-99} benchmarks}
\label{tab:SAT_pramod}
\begin{tabular}{*{14}{c}}
\hline

\multirow{3}{*}{\textbf{Benchmark}}
& \multicolumn{6}{c}{\textbf{10\% Camouflaging Scale}} 
&
& \multicolumn{6}{c}{\textbf{20\% Camouflaging Scale}} \\
\cline{2-7}
\cline{9-14}

& \textbf{\cite{zhang16}} 
& \textbf{\cite{rajendran13_camouflage}} 
& \textbf{\cite{rajendran13_camouflage}}
& \textbf{\cite{nirmala16}}
& \textbf{\cite{wang16_MUX}} 
& \textbf{Our}
&

& \textbf{\cite{zhang16}} 
& \textbf{\cite{rajendran13_camouflage}} 
& \textbf{\cite{rajendran13_camouflage}}
& \textbf{\cite{nirmala16}}
& \textbf{\cite{wang16_MUX}} 
& \textbf{Our}
\\

& (2)$^*$
& (3)$^*$
& (4)$^*$
& (8)$^*$
& (16)$^*$
& (10--14)$^*$
&
& (2)$^*$
& (3)$^*$
& (4)$^*$
& (8)$^*$
& (16)$^*$
& (10--14)$^*$
\\ \hline
 
b14\_C
& 13.77 & 36.12 & 87.14 & 230.05 & 9,735.42 & 136.66
&
& 31.09 & 186.46 & 794.64 & 39,341.2 & 48,399.9 & 6,674.53
\\ %hline

b15\_C
& 20.93 & 77.24 & 172.27 & 1,010.27 & 5,070.97 & 708.31 
&
& 69.21 & 621.75 & 1,592.64 & 8,938.47 & t-o & 4,375.3
\\ %hline

b17\_C
& 256.71 & 1,174.39 & 2,103.08 & 23,408.4 & t-o & 15,807.3 
&
& 864.55 & 11,202.6 & 30,629 & t-o & t-o & t-o
\\ %hline

b20\_C
& 56.35 & 176.81 & 323.81 & 1,909.9 & 8,755.17 & 1,117.02 
&
& 160.731 & 1,839.28 & 22,371.4  & 56,492.6 & t-o & t-o
\\ %hline

b22\_C
& 121.25 & 446.33 & 1,128.02 & 6,289.34 & 25,814.6 & 2,864.64 
&
& 1,291.26 & 10,835.2 & 22,309.8 & t-o & t-o & t-o
\\ %hline

c3540
& 1.47 & 3.32 & 4.08 & 12.04 & 27.72 & 22.02 
&
& 3.58 & 7.88 & 16.10 & 54.24 & 599.95 & 50.80
\\ %hline

c5315
& 0.65 & 1.85 & 4.03 & 12.88 & 22.90 & 7.73 
&
& 1.86 & 7.11 & 20.90 & 144.18 & 369.28 & 44.57
\\ %hline

c7552
& 1.13 & 21.39 & 8.62 & 33.54 & 132.96 & 19.89 
&
& 4.53 & 35.26 & 106.64 & 237.83 & 884.75 & 219.79
\\ \hline

& \multicolumn{6}{c}{\textbf{30\% Camouflaging Scale}} 
&
& \multicolumn{6}{c}{\textbf{40\% Camouflaging Scale and Beyond$^\S$}}  \\
\hline
 
b14\_C
& 115.78 & 1,659.75 & 14,180.7 & t-o & t-o & t-o
&
& 2,560.86 & 96,271.4 & t-o & t-o & t-o & t-o
\\ %hline

b15\_C
& 349.34 & 2,481.25 & 9,165.07 & t-o & t-o & t-o
&
& 759.22 & 24,442.9 & t-o & t-o & t-o & t-o
\\ %hline

b17\_C
& 3,046.33 & 68,028.1 & t-o & t-o & t-o & t-o 
&
& 14,970.7 & t-o & t-o & t-o & t-o & t-o
\\ %hline

b20\_C
& 2,406.48 & 6,027.25 & t-o & t-o & t-o & t-o 
&
& 5,245.08 & 155,278 & t-o & t-o & t-o & t-o
\\ %hline

b22\_C
& 3,754.7 & t-o & t-o & t-o & t-o & t-o 
&
& 16,923.9 & t-o & t-o & t-o & t-o & t-o
\\ %hline

c3540
& 6.30 & 35.25 & 95.48 & 22,613.6 & t-o & t-o 
&
& 26.29 & 747.53 & 3,478.63 & t-o & t-o & t-o
\\ %hline

c5315
& 5.88 & 22.91 & 87.42 & 323.62 & 4,971.03 & 88.92
&
& 13.50 & 62.80 & 1,538.92 & 1,447.68 & t-o & 2,021.74
\\ %hline

c7552
& 26.40 & 91.07 & 1,618.37 & 14,017 & 9,127.99 & 21,850.9 
&
& 48.33--28,448.8$^\S$ & 171.55 & 4,532.72 & t-o & t-o & t-o
\\ \hline

\end{tabular}
\\[1mm]
For a fair evaluation, the
same sets of gates are camouflaged across all camouflaging techniques:
for a given benchmark, gates are randomly selected once and
then memorized. Ten such random sets are generated for each benchmark.
Time-out ``t-o'' is 48 hours, i.e., 172,800 seconds.
$^*$Number of obfuscated functionalities we assume; refer to related publication for the actual sets of functionalities.
$^\S$The runtime range reported for \textit{ISCAS-85} benchmark \textit{c7552} being obfuscated with~\cite{zhang16} is for 40--100\% camouflaging
scale. All other runtimes here correspond to 
40\% camouflaging scale; we found that 50\% camouflaging scale occurs t-o for all other cases.
\end{table*}

\begin{table*}[tb]
\centering
\scriptsize
\setlength{\tabcolsep}{1.95mm}
\caption{Average Runtimes (in Seconds) for the SAT-Based \textit{Double DIP} Attack \cite{shen17,shen17_code} on Selected \textit{ISCAS-85} and \textit{ITC-99} benchmarks}
\label{tab:SAT-double-DIP}
\begin{tabular}{*{14}{c}}
\hline

\multirow{3}{*}{\textbf{Benchmark}}
& \multicolumn{6}{c}{\textbf{10\% Camouflaging Scale}} 
&
& \multicolumn{6}{c}{\textbf{20\% Camouflaging Scale}} \\
\cline{2-7}
\cline{9-14}

& \textbf{\cite{zhang16}} 
& \textbf{\cite{rajendran13_camouflage}} 
& \textbf{\cite{rajendran13_camouflage}}
& \textbf{\cite{nirmala16}}
& \textbf{\cite{wang16_MUX}} 
& \textbf{Our}
&

& \textbf{\cite{zhang16}} 
& \textbf{\cite{rajendran13_camouflage}} 
& \textbf{\cite{rajendran13_camouflage}}
& \textbf{\cite{nirmala16}}
& \textbf{\cite{wang16_MUX}} 
& \textbf{Our}
\\

& (2)$^*$
& (3)$^*$
& (4)$^*$
& (8)$^*$
& (16)$^*$
& (10--14)$^*$
&
& (2)$^*$
& (3)$^*$
& (4)$^*$
& (8)$^*$
& (16)$^*$
& (10--14)$^*$
\\ \hline
 
b14\_C
& 33.37 & 111.33 & 208.44 & 895.31 & 21,105.4 & 401.72
&
& 104.94 & 1,055.62 & 8,321.02 & 57,994.1 & t-o & 40,864.9 
\\ %hline

b15\_C
& 40.43 & 201.85 & 466.89 & 3,175.39 & 8,597.58 & 2,114.4
&
& 180.37 & 1,875.68 & 7,384.14 & 77,413.2 & t-o & 10,978.8 
\\ %hline

b17\_C
& 525.67 & 3,697.23 & 5,981.08 & 42,388.8 & t-o & 28,947.9 
&
& 3,144.69 & 26,809.5 & t-o & t-o & t-o & t-o
\\ %hline

b20\_C
& 136.61 & 481.91 & 867.28 & 7,139.19 & 36,219.7 & 2,428.97
&
& 565.67 & 5,623.08 & t-o & t-o & t-o & t-o
\\ %hline

b22\_C
& 328.95 & 1,190.92 & 5,400.59 & 26,008.8 & t-o & 8,251.29 
&
& 2,741.11 & 18,991.2 & 52,371.2 & t-o & t-o & t-o
\\ %hline

c3540
& 3.46 & 5.66 & 8.18 & 43.10 & 78.38 & 12.27
&
& 7.18 & 14.62 & 64.59 & 363.86 & 41,367.5 & 2,786.15
\\ %hline

c5315
& 1.40 & 5.36 & 12.81 & 30.82 & 51.28 & 18.04
&
& 3.54 & 21.45 & 53.15 & 453.45 & 1,490.21 & 133.41
\\ %hline

c7552
& 2.91 & 7.07 & 12.72 & 93.26 & 248.06 & 32.68
&
& 12.47 & 39.75 & 168.51 & 434.51 & 1,924.91 & 2,902.32
\\ \hline

& \multicolumn{6}{c}{\textbf{30\% Camouflaging Scale}} 
&
& \multicolumn{6}{c}{\textbf{40\% Camouflaging Scale and Beyond$^\S$}}  \\
\hline
 
b14\_C
& 766.24 & 15,403.6 & 70,325.5 & t-o & t-o & t-o
&
& 37,126.4 & t-o & t-o & t-o & t-o & t-o
\\ %hline

b15\_C
& 977.73 & 5,795.87 & 65,201.1 & t-o & t-o & t-o 
&
& 3,182.87 & t-o & t-o & t-o & t-o & t-o 
\\ %hline

b17\_C
& 12,726.7 & t-o & t-o & t-o & t-o & t-o
&
& t-o & t-o & t-o & t-o & t-o & t-o
\\ %hline

b20\_C
& 9,980.07 & 24,536.5 & t-o & t-o & t-o & t-o 
&
& 40,946.3 & t-o & t-o & t-o & t-o & t-o 
\\ %hline

b22\_C
& 10,383.5 & 145,677 & t-o & t-o & t-o & t-o 
&
& t-o & t-o & t-o & t-o & t-o & t-o 
\\ %hline

c3540
& 22.35 & 698.09 & 1,426.13 & t-o & t-o & t-o 
&
& 320.72 & 31,375.4 & t-o & t-o & t-o & t-o 
\\ %hline

c5315
& 11.34 & 40.66 & 240.28 & 918.40 & t-o & 1,879.85
&
& 35.93 & 1,398.75 & 90,724.9 & 46,661.2 & t-o & t-o
\\ %hline

c7552
& 65.49 & 2,205.62 & 22,785.3 & t-o & t-o & t-o 
&
& 137.8--106,213$^\S$ & 25,862.3 & 9,782.61 & t-o & t-o & t-o 
\\ \hline

\end{tabular}
\\[1mm]
Refer to Table~\ref{tab:SAT_pramod} for footnotes.
\end{table*}

\begin{table}[tb]
\centering
\scriptsize
\caption{Runtime, in Seconds, for the SAT-Based Attack \cite{subramanyan15,code_pramod} on Interconnects Obfuscation~\cite{chen18_interconnects} on Selected \textit{ITC-99} benchmarks}
\setlength{\tabcolsep}{1.7mm}
\label{tab:chen18_attack}
\begin{tabular}{*{7}{c}}
\hline

\textbf{Benchmark} 
& \textbf{N1} & \textbf{Time for N1} 
& \textbf{N2} & \textbf{Time for N2} 
& \textbf{N3} & \textbf{Time for N3} 
\\ \hline

b14\_C
& 30 & 7 
& 36 & 9 
& 55 & 11 \\ %hline
b15\_C
& 38 & 7 
& 44 & 8 
& 84 & 15\\ %hline
b17\_C 
& 92 & 149 
& 198  & 170 
& 272 & 214\\ %hline
b18\_C
& 265 & 2,964 
& 334  & 3,223 
& 518 & 3,816\\ %hline
b19\_C 
& 438 & 4,685 
& 583 & 5,393 
& 893 & 7,684\\ %hline
b20\_C 
& 48 & 35 
& 85 & 46 
& 166 & 70\\ %hline
b21\_C 
& 54 & 29 
& 76 & 56 
& 168 & 63\\ %hline
b22\_C 
& 76 & 58 
& 113 & 79 
& 191 & 128\\ \hline
\end{tabular}
\\[1mm]
We model the attack as outlined in Fig.~3 of~\cite{yu17}, i.e., by inserting 2-to-1 MUXes fed by real and dummy wires.
Columns N1, N2, and N3 quote the number of dummy wires,
while limiting the delay overheads to 0\%, 3\%, and 5\%, respectively, as proposed
in~\cite{chen18_interconnects}.
Since the exact locations of dummy wires and other details are not reported in~\cite{chen18_interconnects}, we insert N1--N3 wires randomly into
the netlists available at our end.
Time for N1--N3 denote the total runtime for attacking the related netlists.
\end{table}

\begin{table}[tb]
\centering
\scriptsize
\setlength{\tabcolsep}{0.65mm}
\caption{Results for Proximity Attack \cite{wang2018cat,code_network_flow_attack}
		on Selected \textit{ISCAS-85} Benchmarks}
\label{tab:CCR_results_NFA}
\begin{tabular}{*{10}{c}}
\hline

\multirow{2}{*}{\textbf{Benchmark}}
& \textbf{LC} 
&
& \multicolumn{3}{c}{\textbf{Split after M3}}
&
& \multicolumn{3}{c}{\textbf{Split after M4}}
\\
\cline{4-6}
\cline{8-10}

& \textbf{Scale}
&
& \textbf{CI}
& \textbf{CCR}
& \textbf{R-T}
&
& \textbf{CI}
& \textbf{CCR}
& \textbf{R-T}
\\
\hline 

\multirow{6}{*}{c1908} & Original
&
& 39 & 66 & 0.54 
&
& 11 & 45 & 0.54 \\

& 20\%
&
& 142 & 34 & 0.92 
&
& 33 & 33 & 0.99 \\ 
& 40\%
&
& 208 & 21 & 2.35 
&
& 178 & 19 & 1.23 \\ 
& 60\%
&
& 284 & 15 & 11.59 
&
& 265 & 43 & 3.76 \\ 
& 80\%
&
& 356 & 11 & 17.85 
&
& 327 & 29 & 6.58 \\ 
& 100\%
&
& 411 & 9 & 30.11 
&
& 371 & 25 & 8.49 \\ 
\hline

\multirow{6}{*}{c3540} & Original
&
& 291 & 78 & 2.07 
&
& 22 & 100 & 1.89 \\ 

& 20\%
&
& 778 & 26 & 296.84 
&
& 463 & 33 & 25.68 \\ 
& 40\%
&
& 878 & 20 & 377.87 
&
& 608 & 24 & 44.39 \\ 
& 60\%
&
& 1,017 & 11 & 728.34 
&
& 819 & 19 & 94.71 \\ 
& 80\%
&
& 1,224 & 12 & 1,420.87 
&
& 1,032 & 26 & 206.45 \\ 
& 100\%
&
& 1,409 & 6 & 2,138.83 
&
& 1,262 & 22 & 1,028.66 \\
\hline

\multirow{6}{*}{c5315} & Original
&
& 405 & 55 & 5.21 
&
& 135 & 40 & 2.15 \\ 

& 20\%
&
& 1,002 & 29 & 353.64 
&
& 574 & 25 & 19.02 \\ 
& 40\%
&
& 1,325 & 23 & 840.93 
&
& 930 & 32 & 141.66 \\ 
& 60\%
&
& 1,490 & 13 & 2,696.78 
&
& 1,182 & 28 & 448.9 \\ 
& 80\%
&
& 1,582 & 9 & 3,529.51 
&
& 1,363 & 26 & 1,157.74 \\ 
& 100\%
&
& 1,866 & 10 & 4,857.38 
&
& 1,610 & 28 & 660.14 \\
\hline

\multirow{6}{*}{c7552} & Original
&
& 252 & 48 & 5.02 
&
& 99 & 63 & 5.18 \\ 

& 20\%
&
& 950 & 29 & 495.7 
&
& 594 & 20 & 41.66 \\ 
& 40\%
&
& 1,249 & 18 & 632.97 
&
& 959 & 22 & 265.67 \\ 
&  60\%
&
& 1,489 & 12 & 2,588.73 
&
& 1,294 & 20 & 774.15 \\ 
& 80\%
&
& 1,818 & 10 & 3,525.18 
&
& 1,645 & 26 & 1,515.38 \\ 
& 100\%
&
& 2,011 & 8 & 11,358.27 
&
& 1,891 & 28 & 2,408.11 \\
\hline

\textbf{Avg.\ for Original} & --  
&
& 246.75 & 61.75 & 3.21 
&
& 66.75 & 62 & 2.44 \\
\textbf{Avg.\ for 20\% LC} & -- 
&
& 718 & 29.5 & 286.78 
&
& 416 & 27.75 & 21.84 \\
\textbf{Avg.\ for 40\% LC} & -- 
&
& 915 & 20.5 & 463.52 
&
& 668.75 & 24.25 & 113.24 \\
\textbf{Avg.\ for 60\% LC} & -- 
&
& 1,070 & 12.75 & 1,506.36 
&
& 890 & 27.5 & 330.38 \\
\textbf{Avg.\ for 80\% LC} & -- 
&
& 1,245 & 10.5 & 2,123.35 
&
& 1,091.75 & 26.75 & 721.54 \\
\textbf{Avg.\ for 100\% LC} & --
&
& 1,424.25 & 8.25 & 4,596.15 
&
& 1,283.5 & 25.75 & 1.026.35 \\
\hline
\end{tabular}
\\[1mm]
CCR is correct connection rate (\%), CI is cut inputs, and R-T is runtime (s).
\end{table}

Regarding the impact of the camouflaging scale and split layer, we observe the following.
First, for both layers,
there is a significant CCR reduction already for 20\% camouflaging over original layouts, namely by $\approx$2.1$\times$.
This implies that
a relatively small-scaled
application of our scheme is sufficient to significantly weaken this
attack~\cite{wang2018cat,code_network_flow_attack}, and this holds true across split layers.
Second, for M3, increasing the camouflaging scale helps to reduce the CCR notably further, whereas the CCR remains at $\approx$26.4\% for M4 across camouflaging scales.
This implies that our scheme is more effective when lower split layers are considered, which can be expected and has been discussed before in general, e.g., in~\cite{sengupta17_SM_ICCAD}.

We also study the
quality (or rather, lack thereof)
for the netlists recovered by~\cite{wang2018cat,code_network_flow_attack}.
In Table~\ref{tab:NFA_results_OER_HD}, we report HD and OER values across camouflaging scales and split layers.
The OER values reveal errors for all the netlists, irrespective of the camouflaging scale. Concerning HD, we observe many
errors as well, already for
20\% camouflaging scale, and for full-chip camouflaging, HD is approaching the ideal value (50\%) for all the netlists recovered
across both split layers.
Therefore, aside from CCR, we also demonstrate that netlists recovered by~\cite{wang2018cat,code_network_flow_attack}
deviate significantly in terms of functional behavior from the original layouts.

\begin{table}[tb]
\centering
\scriptsize
\setlength{\tabcolsep}{0.6mm}
\caption{HD, OER for Netlists Obtained by Proximity Attack \cite{wang2018cat,code_network_flow_attack}}
\label{tab:NFA_results_OER_HD}
\begin{tabular}{*{16}{c}}
\hline

\textbf{LC Scale:}
&
& \multicolumn{2}{c}
{\textbf{20\%}} 
&
& \multicolumn{2}{c}
{\textbf{40\%}}
&
& \multicolumn{2}{c}
{\textbf{60\%}}
&
& \multicolumn{2}{c}
{\textbf{80\%}}
&
& \multicolumn{2}{c}
{\textbf{100\%}}
\\
\cline{1-1}
\cline{3-4}
\cline{6-7}
\cline{9-10}
\cline{12-13}
\cline{15-16}

\multirow{2}{*}{\textbf{Benchmark}}
&
& \textbf{HD} & \textbf{OER}
&
& \textbf{HD} & \textbf{OER}
&
& \textbf{HD} & \textbf{OER}
&
& \textbf{HD} & \textbf{OER}
&
& \textbf{HD} & \textbf{OER}
\\
\cline{3-16}

\multicolumn{16}{c}{\textbf{Split after M3}} \\
\hline

c1908 & 
&
38.93 & 99.99 & 
&
40.14 & 99.99 & 
&
45.67 & 99.99 &
&
47.71 & 99.99 & 
&
49.85 & 99.99  
\\ %hline

c3540 & 
&
42.74 & 99.99 & 
&
44.59 & 99.99 & 
&
45.86 & 99.99 &
&
49.67 & 99.99 & 
&
49.79 & 99.99  
\\ %hline

c5315 & 
&
43.23 & 99.99 & 
&
46.81 & 99.99 & 
&
47.89 & 99.99 &
&
49.51 & 99.99 & 
&
50.78 & 99.99  
\\ %hline

c7552 & 
&
45.62 & 99.99 & 
&
47.5 & 99.99 & 
&
49.04 & 99.99 &
&
47.78 & 99.99 & 
&
48.73 & 99.99  
\\ \hline

\textbf{Average} & 
&
42.63 & 99.99 & 
&
44.76 & 99.99 & 
&
47.11 & 99.99 &
&
48.67 & 99.99 & 
&
49.79 & 99.99  
\\ \hline

\multicolumn{16}{c}{\textbf{Split after M4}} \\
\hline

c1908 & 
&
35.49 & 99.99 & 
&
38.34 & 99.99 & 
&
43.89 & 99.99 &
&
46.31 & 99.99 & 
&
47.43 & 99.99  
\\ %hline

c3540 & 
&
40.34 & 99.99 & 
&
43.19 & 99.99 & 
&
44.56 & 99.99 &
&
47.91 & 99.99 & 
&
48.43 & 99.99  
\\ %hline

c5315 & 
&
42.87 & 99.99 & 
&
43.87 & 99.99 & 
&
45.91 & 99.99 &
&
47.17 & 99.99 & 
&
48.78 & 99.99  
\\ %hline

c7552 & 
&
44.12 & 99.99 & 
&
45.52 & 99.99 & 
&
46.78 & 99.99 &
&
45.78 & 99.99 & 
&
47.89 & 99.99  
\\ \hline

\textbf{Average} & 
&
40.71 & 99.99 & 
&
42.73 & 99.99 & 
&
45.29 & 99.99 &
&
46.79 & 99.99 & 
&
48.13 & 99.99  
\\ \hline
\end{tabular}
\\[1mm]
HD and OER are Hamming distance and output error rates, respectively; both are reported in percentage.
\end{table}

Finally, we also conduct the routing-congestion-aware attack \textit{crouting}
by Maga\~{n}a \textit{et al.}~\cite{magana17,code_MAGANA_attack} for the larger
\textit{ITC-99} benchmarks (Table~\ref{tab:crouting}).
We observe a significant increase of \textit{vpins} once larger camouflaging scales are effected:
across all benchmarks, on average, the increase is 12.55$\times$ for 100\% camouflaging when compared to original, unprotected layouts.
The number of candidates \textit{E[LS]}, which represents the possible pairings for each cut net to consider, increases accordingly as well.
As a result, the attack complexity, represented by \textit{FOM},
increases on average by 9.30$\times$ and 24.15$\times$ for split layers M3 and M4, respectively.
Considering the impact of the camouflaging scale and split layer,
we observe similar trends as above. That is, we note already for 20\% camouflaging a significant increase in the \textit{FOM}, for both 
split layers. Furthermore, our scheme remains more resilient at M3, while \textit{FOM} for M4
increase well with the camouflaging scale here.
Overall, we find that our scheme weakens the prospects of
the \textit{crouting} attack~\cite{magana17,code_MAGANA_attack} significantly.

\begin{table}[tb]
\centering
\scriptsize
\setlength{\tabcolsep}{1.3mm}
\caption{Results for \textit{crouting} Proximity Attack~\cite{magana17,code_MAGANA_attack} on Selected \textit{ITC-99} Benchmarks}
\label{tab:crouting}
\begin{tabular}{*{11}{c}}
\hline

\multirow{2}{*}{\textbf{Benchmark}}
&
& \textbf{LC}
&
& \multicolumn{3}{c}{\textbf{Split after M3}}
&
& \multicolumn{3}{c}{\textbf{Split after M4}}
\\
\cline{5-7}
\cline{9-11}

&
& \textbf{Scale}
&
& \textbf{\textit{vpins}}
& \textbf{\textit{E[LS]}}
& \textbf{\textit{FOM}}
&
& \textbf{\textit{vpins}}
& \textbf{\textit{E[LS]}}
& \textbf{\textit{FOM}}
\\ \hline

\multirow{6}{*}{b14\_C}
&
& Original & 
&
224 & 1.78 & 0.04 &
&
28 & 0.32 & 0.01
\\ %\cline{2-8}
&
& 20\% & 
&
1,146 & 13.03 & 0.27 &
&
950 & 12.36 & 0.26
\\ %\cline{2-8}
&
& 40\% & 
&
1,832 & 18.26 & 0.38 &
&
1,624 & 17.1 & 0.36
\\ %\cline{2-8}
&
& 60\% & 
&
2,234 & 19.78 & 0.41 &
&
2,062 & 18.98 & 0.4
\\ %\cline{2-8}
&
& 80\% & 
&
3,060 & 21.01 & 0.44 &
&
2,924 & 20.31 & 0.42
\\ %\cline{2-8}
&
& 100\% & 
&
3,310 & 22.02 & 0.46 &
&
3,144 & 21.29 & 0.44
\\ \hline

\multirow{6}{*}{b15\_C}
&
& Original & 
&
458 & 2.74 & 0.06 &
&
174 & 1.59 & 0.03
\\ %\cline{2-8}
&
& 20\% & 
&
1,342 & 9.88 & 0.27 &
&
1,144 & 10.44 & 0.29
\\ %\cline{2-8}
&
& 40\% & 
&
2,060 & 12.37 & 0.34 &
&
1,930 & 12.71 & 0.35
\\ %\cline{2-8}
&
& 60\% & 
&
2,852 & 17.82 & 0.37 &
&
2,820 & 18.38 & 0.38
\\ %\cline{2-8}
&
& 80\% & 
&
3,430 & 18.54 & 0.39 &
&
3,478 & 19.17 & 0.4
\\ %\cline{2-8}
&
& 100\% & 
&
3,988 & 19.95 & 0.42 &
&
3,972 & 20.33 & 0.42
\\ \hline

\multirow{6}{*}{b17\_C}
&
& Original & 
&
1,500 & 9.08 & 0.05 &
&
68 & 4.42 & 0.02
\\ %\cline{2-8}
&
& 20\% & 
&
4,382 & 36.78 & 0.23 &
&
3,310 & 40.65 & 0.26
\\ %\cline{2-8}
&
& 40\% & 
&
6,628 & 38.06 & 0.28 &
&
N/A & N/A & N/A
\\ %\cline{2-8}
&
& 60\% & 
&
9,050 & 53.19 & 0.28 &
&
8,724 & 53.27 & 0.28
\\ %\cline{2-8}
&
& 80\% & 
&
11,258 & 59.44 & 0.3 &
&
11,230 & 60.15 & 0.31
\\ %\cline{2-8}
&
& 100\% & 
&
12,586 & 65.3 & 0.33 &
&
12,234 & 63.69 & 0.33
\\ \hline

\multirow{6}{*}{b20\_C}
&
& Original & 
&
448 & 3.21 & 0.03 &
&
100 & 0.82 & 0.01
\\ %\cline{2-8}
&
& 20\% & 
&
2,714 & 28.19 & 0.3 &
&
2,068 & 27.72 & 0.3
\\ %\cline{2-8}
&
& 40\% & 
&
4,116 & 35.18 & 0.37 &
&
3,600 & 33.17 & 0.35
\\ %\cline{2-8}
&
& 60\% & 
&
5,386 & 24.45 & 0.39 &
&
5,182 & 23.9 & 0.38
\\ %\cline{2-8}
&
& 80\% & 
&
6,616 & 27.39 & 0.43 &
&
6,378 & 26.73 & 0.42
\\ %\cline{2-8}
&
& 100\% & 
&
7,880 & 51.61 & 0.43 &
&
7,438 & 48.8 & 0.41
\\ \hline

\multirow{6}{*}{b21\_C}
&
& Original & 
&
520 & 3.58 & 0.04 &
&
106 & 0.96 & 0.01
\\ %\cline{2-8}
&
& 20\% & 
&
2,636 & 24.26 & 0.31 &
&
1,942 & 22.41 & 0.29
\\ %\cline{2-8}
&
& 40\% & 
&
3,826 & 32.64 & 0.34 &
&
3,366 & 31.29 & 0.33
\\ %\cline{2-8}
&
& 60\% & 
&
4,938 & 36.3 & 0.38 &
&
4,702 & 35.88 & 0.38
\\ %\cline{2-8}
&
& 80\% & 
&
7,026 & 39.98 & 0.4 &
&
6,852 & 37.86 & 0.38
\\ %\cline{2-8}
&
& 100\% & 
&
7,910 & 43.87 & 0.44 &
&
7,400 & 41.31 & 0.41
\\ \hline

\multirow{6}{*}{b22\_C}
&
& Original & 
&
974 & 4.85 & 0.05 &
&
312 & 1.86 & 0.02
\\ %\cline{2-8}
&
& 20\% & 
&
3,966 & 32.87 & 0.28 &
&
2,978 & 31.51 & 0.28
\\ %\cline{2-8}
&
& 40\% & 
&
5,886 & 41.11 & 0.35 &
&
5,140 & 39.29 & 0.34
\\ %\cline{2-8}
&
& 60\% & 
&
8,152 & 36.69 & 0.37 &
&
7,634 & 34.99 & 0.35
\\ %\cline{2-8}
&
& 80\% & 
&
9,774 & 40.77 & 0.41 &
&
9,538 & 39.52 & 0.4
\\ %\cline{2-8}
&
& 100\% & 
&
10,362 & 42.57 & 0.43 &
&
9,964 & 41.03 & 0.41
\\ \hline

\end{tabular}
\\[1mm]
The metric \textit{vpin} is number of open two-pin nets, \textit{E[LS]} is the number of candidates over a specific region, and \textit{FOM} is the
figure of merit for attack complexity, as defined in~\cite{magana17} and explained in Sec.~\ref{sec:setup}.
N/A denotes attack failures.
For a fair comparison across different benchmarks and layouts, we setup up each 
run with a bounding-box size equal to
$1/8$ of the half perimeter of the respective die outlines.
\end{table}

\section{Conclusion}
\label{sec:conclusion}

Given the manifold limitations of the prior art for camouflaging---regarding applicability, layout overheads, as well as 
resilience---we
argue that new avenues are called for.
Note that this argument also applies to \textit{provably secure} schemes, where important fallacies have been
	demonstrated, namely algorithmic attacks tailored for identification and/or
	removal of the additional, dedicated circuitry required by such schemes, as well as
	their inherently low output corruptibility.
Here,
we promote the obfuscation of interconnects as such a promising avenue going beyond the prior art,
especially for efficient IP protection at large scales.

Toward this end, we propose and implement
BEOL-centric camouflaging primitives which are applicable to any FEOL node, and integrate them within a commercial-grade CAD framework.
We strive for practically relevant layout evaluation---all our camouflaged layouts are DRC-clean at the GDSII level, and we consider
traditional but also modern, large-scale benchmark suites.
We thoroughly contrast ours to the prior art, both in terms of layout cost and resilience.
For the latter, we have leveraged powerful SAT-based attacks and proximity attacks.
Our scheme hinders both threats ultimately by virtue of scale and entanglement.
For SAT-based attacks, we show that tackling our large-scale camouflaging scheme becomes
computationally too expensive; for proximity attacks, we show that the large-scale increase of cut wires at the FEOL-BEOL interface, which our
schemes incurs by construction, is reducing the effectiveness of such attacks.

To the best of our knowledge, the proposed scheme is the first that can deliver low-cost \textit{and} resilient full-chip camouflaging. That is
especially true when also considering the protection offered against fab adversaries, enabled by split manufacturing applied in conjunction with
our notion of obfuscating the interconnects.

As for future work, we will study in more detail the formal underpinnings for computational cost induced on algorithmic attacks by our scheme.
We also plan to demonstrate that our scheme suffers neither from low output corruptibility, rendering it resilient against approximate SAT-based
attacks like \textit{AppSAT}~\cite{shamsi17}, nor from removal
attacks~\cite{bypass-attack2017,yasin17_TETC}, which is both in contrast to provably secure schemes.
A straightforward intuition here is that our scheme intertwines camouflaging within the netlist at large scales,
where any false inferences can propagate throughout the whole netlist, whereas provably secure schemes
protect only particular patterns/outputs.

\def\bibfont{\footnotesize}
% Generated by IEEEtran.bst, version: 1.14 (2015/08/26)

\begin{IEEEbiography}[{\includegraphics[width=1in,height=1.25in,clip,keepaspectratio]
{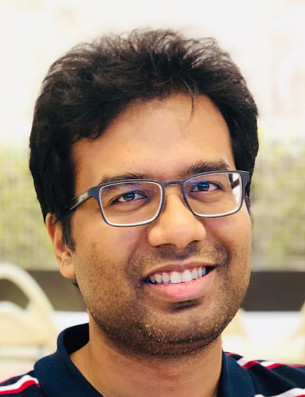}}]{Satwik Patnaik} (S'16)
received B.E.\
in Electronics and Telecommunications from the University of
Pune, India and
M.Tech.\ in Computer Science and Engineering with a specialization in VLSI Design from Indian Institute of Information Technology and
Management, Gwalior, India. 
He is a final year Ph.D.\ candidate at the Department of Electrical and Computer Engineering at the 
Tandon School of Engineering with New York University, Brooklyn, 
NY, USA.
He is
also a Global Ph.D.\ Fellow with New York University Abu Dhabi, U.A.E.
He received the Bronze Medal in the Graduate category at the ACM/SIGDA Student Research Competition (SRC) held at ICCAD 2018, and the best paper award at the Applied Research Competition (ARC) held in conjunction with Cyber Security Awareness Week (CSAW), 2017.
His current research interests 
include Hardware security, Trust and reliability issues for CMOS and emerging devices 
with particular focus on low-power VLSI Design.
He is a student member of IEEE and ACM.
\end{IEEEbiography}

\begin{IEEEbiography}[{\includegraphics[width=1in,height=1.25in,clip,keepaspectratio]{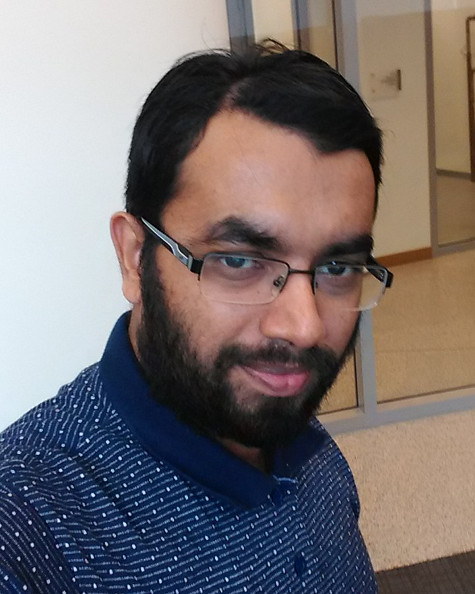}}]{Mohammed Ashraf}
is a 
Senior Physical Design engineer from India. 
He obtained his Bachelor's degree in electronics and 
telecommunication engineering from College of Engineering Trivandrum, Kerala, in 2005. 
He carries an experience of 10 years in the VLSI industry. 
He has worked with various multi-national companies like NVIDIA Graphics, Advanced Micro Devices (AMD), and Wipro Technologies. 
He worked also with Dubai Circuit Design, Dubai Silicon Oasis, UAE.
Mr.\ Ashraf is currently a Research Engineer at Center for Cyber Security (CCS) at New York University Abu Dhabi.
\end{IEEEbiography}

\begin{IEEEbiography}[{\includegraphics[width=1in,height=1.25in,clip,keepaspectratio]{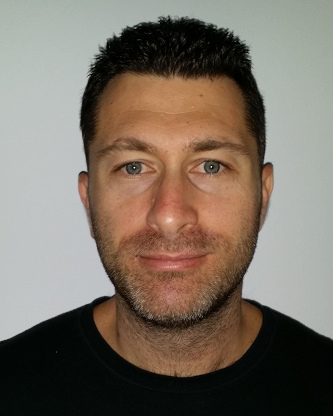}}]{Ozgur Sinanoglu} (M'11--SM'15)
is a Professor of Electrical and Computer Engineering at New York University Abu Dhabi. He earned his B.S.\ degrees, one in
Electrical and Electronics Engineering and one in Computer Engineering, both from Bogazici University, Turkey in 1999. He obtained his MS
and PhD in Computer Science and Engineering from University of California San Diego in 2001 and 2004, respectively. He has industry
experience at TI, IBM and Qualcomm, and has been with NYU Abu Dhabi since 2010. During his PhD, he won the IBM PhD fellowship award twice.
He is also the recipient of the best paper awards at IEEE VLSI Test Symposium 2011 and ACM Conference on Computer and Communication Security
2013. 

Prof.\ Sinanoglu's research interests include design-for-test, design-for-security and design-for-trust for VLSI circuits, where he has more
than 180 conference and journal papers, and 20 issued and pending US Patents. Prof.\ Sinanoglu has given more than a dozen tutorials on hardware
security and trust in leading CAD and test conferences, such as DAC, DATE, ITC, VTS, ETS, ICCD, ISQED, etc. He is serving as track/topic
chair or technical program committee member in about 15 conferences, and as (guest) associate editor for IEEE TIFS, IEEE TCAD, ACM JETC,
      IEEE TETC, Elsevier MEJ, JETTA, and IET CDT journals. 

Prof.\ Sinanoglu is the director of the Design-for-Excellence Lab at NYU Abu Dhabi. His recent research in hardware security and trust
is being funded by US National Science Foundation, US Department of Defense, Semiconductor Research Corporation, Intel Corp and
Mubadala Technology.
\end{IEEEbiography}

\begin{IEEEbiography}[{\includegraphics[width=1in,height=1.25in,clip,keepaspectratio]{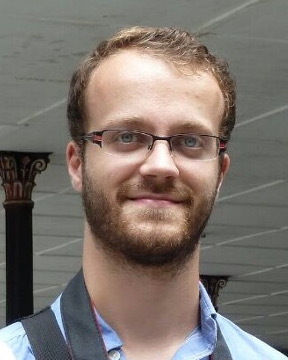}}]{Johann Knechtel}
(M'11)
received the M.Sc.\ in Information Systems Engineering (Dipl.-Ing.) in 2010 and the Ph.D.\ in Computer Engineering
(Dr.-Ing., summa cum laude) in 2014, both from TU Dresden, Germany.  He is a Research Scientist
at the New York University, Abu Dhabi, UAE.  Dr.\ Knechtel was a
Postdoctoral Researcher in 2015--16 at the Masdar Institute of Science and Technology, Abu Dhabi.  From 2010 to 2014, he was
a Ph.D.\ Scholar with the DFG Graduate School on ``Nano- and Biotechnologies for Packaging of Electronic
Systems''
hosted at the TU Dresden.  In 2012, he was a
Research Assistant with the Dept.\ of Computer Science and Engineering, Chinese University of Hong Kong, China.  In 2010, he
was a Visiting Research Student with the Dept.\ of Electrical Engineering and Computer Science, University of Michigan, USA.
His research interests cover VLSI Physical Design Automation, with particular focus on Emerging Technologies and Hardware Security.
\end{IEEEbiography}

\end{document}